\documentclass[namedreferences]{solarphysics}

\usepackage[urlcolor=blue,colorlinks=true,linkcolor=red,citecolor=magenta,breaklinks=true]{hyperref}
\usepackage[optionalrh,natbib]{spr-sola-addons} 
\usepackage{graphicx}        
\usepackage{color}           
\usepackage{breakurl}        
\usepackage{rotating}        
\usepackage{longtable}
\usepackage{enumerate}

\graphicspath{{figs/}}

\newcommand{\adsurl}[1]{\href{http://adsabs.harvard.edu/abs/#1}{ADS}}
\newcommand{\doiurl}[1]{\href{http://dx.doi.org/#1}{DOI}}

\newcommand{\etal}{{\it et al.}}

\newcommand{\eg}{e.g. {\ }}
\newcommand{\ie}{i.e. {\ }}

\newcommand{\fov}{field of view{\ }}
\newcommand{\fovnospace}{field of view}

\newcommand{\degr}{{^\circ}}
\renewcommand{\deg}{$^\circ$}
\newcommand{\Rsun}{\,R$_\odot$}
\newcommand{\Bsun}{$\overline{B_\odot}$} 

\begin{document}

\begin{article}

\begin{opening}

\title{Restoration of the K and F Components of the Solar Corona from LASCO-C2 Images over 24 Years [1996\,--\,2019]}
 
\author[addressref={aff1},email={a.llebaria@laposte.net}]{\inits{A. }\fnm{Antoine}~\lnm{Llebaria}} 	
\author[addressref={aff2},corref,email={philippe.lamy@latmos.ipsl.fr}]{\inits{P. }\fnm{Philippe}~\lnm{Lamy}\orcid{0000-0002-2104-2782}}
\author[addressref={aff2},email={hugo.gilardy@latmos.ipsl.fr}]{\inits{H. }\fnm{Hugo}~\lnm{Gilardy}} 	
\author[addressref={aff1},email={Bricexp@hotmail.com}]{\inits{B. }\fnm{Brice}~\lnm{Boclet}} 	
\author[addressref={aff1},email={jean.loirat@gmail.com}]{\inits{J. }\fnm{Jean}~\lnm{Loirat}} 	 	

\address[id=aff1]{Laboratoire d'Astrophysique de Marseille, CNRS \& Aix-Marseille Universit\'e, 38 rue Fr\'ed\'eric Joliot-Curie, 13388 Marseille cedex 13, France}
\address[id=aff2]{Laboratoire Atmosph\`eres, Milieux et Observations Spatiales, CNRS \& UVSQ, 11 Bd d'Alembert, 78280 Guyancourt, France}
      
\runningauthor{A. Llebaria et al.}
\runningtitle{F-Corona Restoration}
 
\begin{abstract}
We present a photometrically accurate restoration of the K- and F-coronae from white-light images obtained over 24 years [1996\,--\,2019] by the {\it Large-Angle Spectrometric COronagraph} (LASCO-C2) onboard the {\it Solar and Heliospheric Observatory} (SOHO).
The procedure starts with the data set of unpolarized images of 512 $\times$ 512 pixels produced by the polarimetric analysis of the routine C2 polarization sequences 
(Lamy et al. {\it Solar Phys.} {\bf 295}, 89, 2020) in which the F-corona, the instrumental stray light, and possible remnants of the K-corona due to the imperfect polarimetric separation are entangled. 
Disentangling these components requires a complex procedure organized in three stages, each composed of several steps. 
Stage 1 establishes the distinct variations of the radiance of these components with the Sun--SOHO distance, and generate a new data set of median images calculated for each Carrington rotation. 
Stage 2 achieves the restoration of a set of 36 stray-light images that account for the temporal variation of the stray-light pattern, in particular those associated with the periodic roll maneuvers of SOHO which started in 2003.
Stage 3 achieves the restoration of the F-corona, and a time series of daily images is generated.
Combining these images with the set of stray-light images allowed us to process the whole set of routine LASCO-C2 images of 1024 $\times$ 1024 pixels (approximately 626,000 images) and to produce calibrated, high-resolution images of the K-corona.
The two sets of images of the K-corona, that produced by polarimetric separation of 512 $\times$ 512 pixels images and that presently produced by subtraction, are in excellent photometric agreement.
We extend our past conclusions that the temporal variation of the integrated radiance of the K-corona tracks the solar activity over Solar Cycles 23 and 24, and that it is highly correlated with the temporal variation of the total magnetic field.
The behaviours of the integrated radiance during the last few years of the declining phases of Solar Cycles 23 and 24 are remarkably similar, reaching the same base level and leading to a duration of 11.0 years for the latter cycle, in agreement with that derived from sunspots.

\end{abstract}

\keywords{Corona, F-corona, K-corona, Stray light}
\end{opening}

\section{Introduction}
Most of the analyses published so far of the white-light images obtained with the {\it Large-Angle Spectrometric COronagraph} (LASCO-C2: Brueckner et al., 1995) of the {\it Solar and Heliospheric Observatory} (SOHO: Domingo, Fleck, and Poland, 1995) relied on the subtraction of simple empirical background images to reveal the dynamic features of the K-corona such as coronal mass ejections (CMEs).
These background images are constructed by taking the minimum of the daily median images over a period of typically one month so that the daily median eliminates the short-term changes and the monthly minimum eliminates the longer-term changes.
This technique, as well as others such as running- or base-differences, removes the static component of the images, which comprises the F-corona, the stable background K-corona, and any instrumental stray light.
It was however applied by \cite{Battams2020} to build a coronal brightness index, but the authors admitted that it does remove some amount of K-corona signal ``primarily in persistent streamers in equatorial latitudes and most noticeably around periods of low solar activity''.
In summary, these techniques are appropriate to the qualitative and even quantitative analysis of dynamic structures (\eg CMEs), but questionable for analysis that require accurate quantitative determinations of the F-corona and the global K-corona, which both requires a rigorous restoration.

The very first attempt to build a time series of K-corona images (from March 1996 to January 1999) was performed by \cite{Llebaria1999} for the purpose of studying coronal activity.
They did not elaborate their procedure in their short article, but underlined the complexity of separating the three components present in the LASCO-C2 images: K-corona, F-corona, and stray light.
They further pointed out the importance of taking into account the eccentricity of the orbit of the SOHO spacecraft as well as its inclination with respect to the plane of symmetry of the zodiacal cloud crossed twice per year by SOHO.

The difficulties are well illustrated by the work of \cite{Hayes2001} who endeavored to derive the electron density from the inversion of LASCO-C2 total brightness images.
Their attempt to subtract the F-corona by taking a pixel-by-pixel minimum over a time interval of 56 days failed, and they recognized that their background model did contain some K-corona signal, thus confirming our above reservations about this technique. 
They then imposed an ad-hoc slope on their background model, adjusting it so as to minimize the K-corona contribution at the lower heights, and they derived an approximation of the F-corona.
This highly empirical method strongly depends upon the specific image being analyzed and is not really adapted to the processing of long time series of images.
It was later applied by \cite{Fainshtein2007,Fainshtein2009} to a few C2 images spanning Solar Cycle (hereafter abbreviated to SC) 23 assuming that the slope determined by \cite{Hayes2001} holds throughout this cycle, and further extended it to C3 images using simple equatorial and polar profiles of the electron density to calculate the K-corona to be subtracted. 

Two subsequent articles were devoted to this question, both making use of the method of \cite{Saito1977} implemented to investigate the equatorial and polar K- and F-coronal components on the basis of {\it Skylab} images obtained near the minimum of SC 20. 
Briefly, this method consists in inverting polarized brightness images $pB$ to retrieve the electron density assuming a spherically symmetric spatial distribution, calculating the Thomson scattering, and subtracting the so-generated K-corona component to retrieve the F-component.
Incidentally, a close variant of this method was proposed well before, namely by \cite{Kluber1958}, and also used in the {\it Skylab} context by \cite{Munro1977}.
\cite{Morgan2007} applied this method to four cases of LASCO-C2 images, two during the minimum and two during the maximum of SC 23 and produced equatorial and polar profiles that led them to reach conclusions about the stability of the visible F-corona.
\cite{Dolei2016} considered only one case in March 2008 during the SC 23/24 minimum and produced an intensity map of the F-corona highly distorted by the presence of a bright streamer on the eastern side, complemented by equatorial and polar profiles.
In addition to the over-simplification of a spherical K-corona and the unproven accuracy of their photopolarimetric analysis, both articles ignored the question of the stray light and of the influence of the orbit of SOHO.

\begin{figure}[htpb!]
\begin{center}
\includegraphics[width=\textwidth]{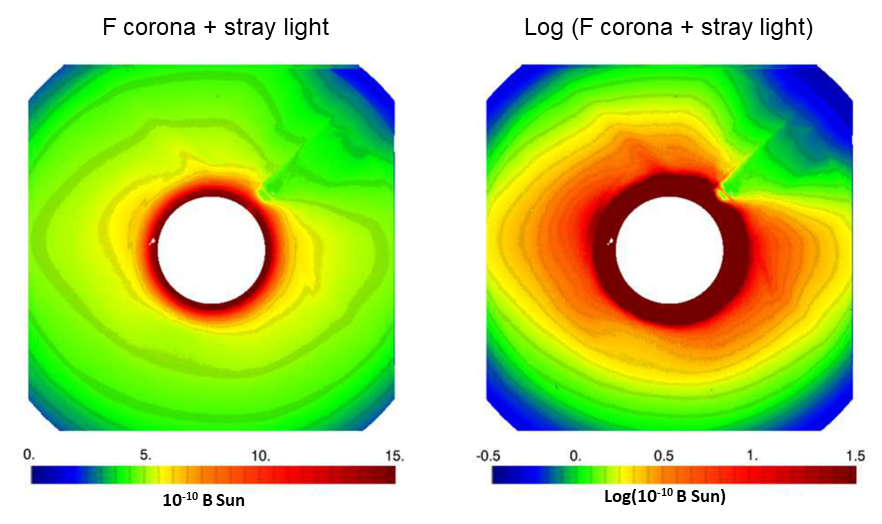}
\caption{Two displays of an image resulting from the polarimetric analysis of \cite{Lamy2020} where the F-corona and the stray light are entangled.
They have not been corrected for the vignetting effect to best reveal the different components of the stray light: the diffraction fringe surrounding the occulter (which extends to 2.2\,\Rsun), several arcs, and diffraction artifacts from the pylon.
The left display uses a linear scale to best view the diffraction fringe.
The right display uses a logarithmic scale to best view the distorted F-corona and the stray-light structures.
For both displays, the units is $10^{-10}$\,\Bsun $\space$.}
\label{fig:F_SL}
\end{center}
\end{figure}

The first thorough analysis of the whole set of polarization sequences of LASCO-C2 images of 512$\times$512 pixels obtained with the orange filter over 24 years [1996\,--\,2019] was performed by \cite{Lamy2020} who showed that extensive corrections are required to produce valid polarization and polarized radiance measurements.
They concentrated on the production of complete sets of K-corona and $pB$ images and inverted the latter images using a two-dimensional procedure developed by \cite{Quemerais2002} to produce maps of the electron density over 24 years [1996\,--\,2019].
They left open the question of separating the two unpolarized components, ``F'' and ``SL'' standing for F-corona and stray light, respectively (Figure~\ref{fig:F_SL}), as well as that of retrieving the K-corona from the more numerous, routine high resolution images of 1024 $\times$ 1024 pixels.
It is the purpose of the present article to tackle these two questions and thus complete the quantitative analysis of the whole set of LASCO-C2 images obtained with the orange filter. 
It is organized as follows:
We first discuss the constraints and obstacles facing the restoration of the F-corona.
We then describe our elaborated procedure, which leads first to the restoration of the stay light and in turn to the restoration of the F-corona from the ``F+SL'' images produced by the polarimetric analysis.
These results are finally employed to extract the K-corona from the routine high-resolution LASCO-C2 images, and we characterize its temporal evolution in comparison with an index and a proxy of solar activity.
We conclude by emphasizing the importance of polarization observations for the accurate restoration of the K- and F-coronae beyond $\approx$2\,R$_\odot$ from images obtained with space coronagraphs.

\section{Constraints and Obstacles Facing the Restoration of the F-corona}
     \label{Sec:Const} 

The photometrically accurate restoration of the F-corona from LASCO-C2 images is by no means a straightforward operation.
We first summarize below the constraints and obstacles coming from the operations of SOHO and of LASCO, the difficulties inherent to the characterization of the F-corona, and the additional complications resulting from the presence of stray light, prominently the diffraction fringe surrounding the occulter.

\subsection{Constraints from SOHO and LASCO Operations}
     \label{Sec:Oper} 

\begin{figure}[htpb!]
\begin{center}
\includegraphics[width=\textwidth]{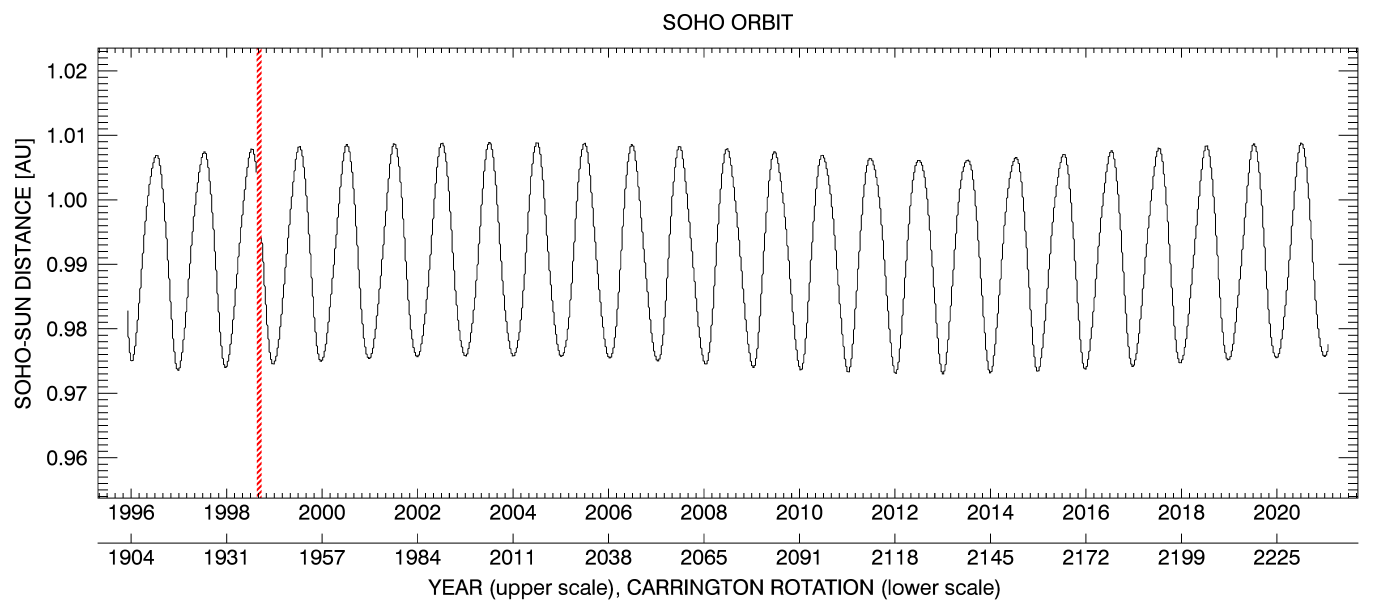}
\caption{Temporal variation of the Sun--SOHO distance over 25 Years [1996\,--\,2020].}
\label{fig:dist}
\end{center}
\end{figure}

SOHO was launched on 2 December 1995 and reached the L${}_1$ Lagrangian point of the Sun--Earth system on 14 February 1996, where it was placed in a halo orbit with a period of $176.313$ days.  
The orbit of SOHO, which is confined to the ecliptic plane, therefore results from a combination of an Earth synchronous elliptic orbit with a period of $365.249$ days and this halo orbit whose period is very close to half of the main orbital period, \ie its second harmonic.  
Consequently, the Sun--SOHO distance oscillates between typically 0.973 and 1.009 AU ($\approx$209 and 217\Rsun) as illustrated in Figure~\ref{fig:dist}.

The operation of SOHO experienced two major interruptions since 1996: first, an accidental loss of control during a roll maneuver on 25 June 1998, which resulted in a long data gap until recovery on 22 October 1998 and second a failure of the gyroscopes which caused another interruption from 21 December 1998 to 6 February 1999 when nominal operation resumed. 
On a short-term basis, there have been minor interruptions of a few days, typically two to three, which have no impact on the present work. 

\begin{figure}[htpb!]
\begin{center}
\includegraphics[width=\textwidth]{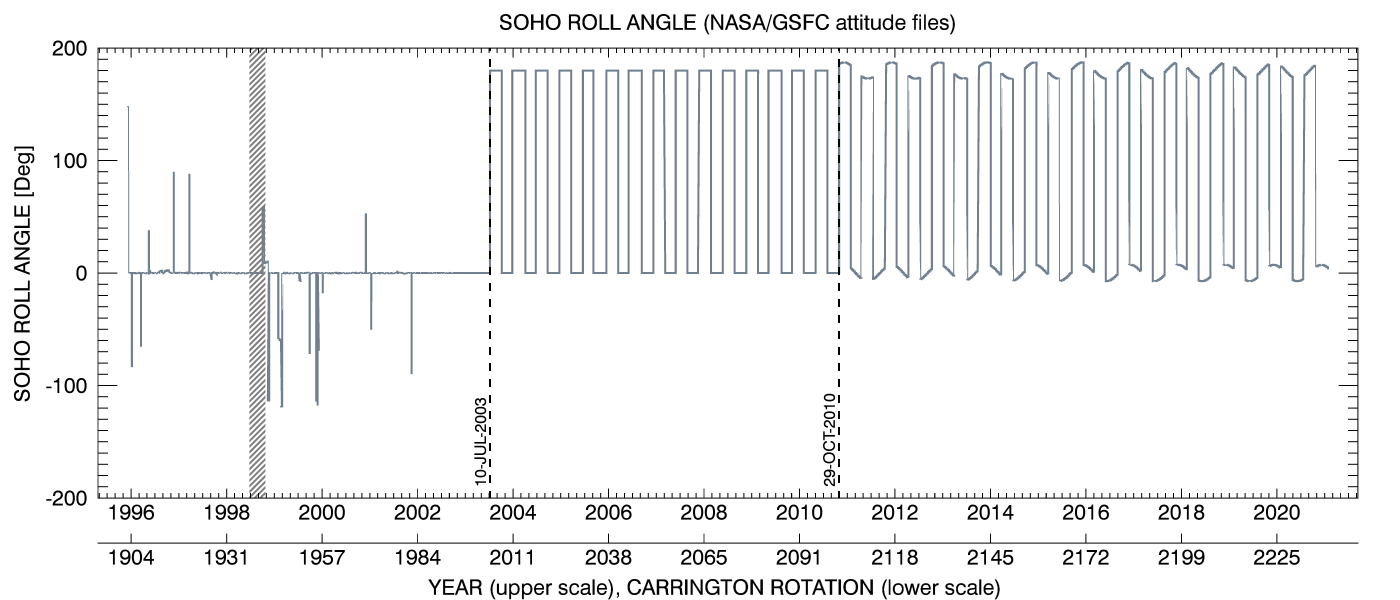}
\caption{Temporal variation of the roll angle of SOHO over 25 Years [1996\,--\,2020].}
\label{fig:roll}
\end{center}
\end{figure}

During its first year of operation, the attitude of SOHO was set such that its reference axis was aligned along the sky-projected direction of the solar rotational axis resulting in this direction being ``vertical'' (\ie along the Y-axis of the CCD detector) with solar North up on the LASCO images.
Starting on 10 July 2003 and following the failure of the motor steering its antenna, SOHO is periodically (every three months) rolled by 180$\degr$ to maximize telemetry transmission to Earth.
As seen on the LASCO images in the instrument reference frame, the sky-projected direction of the solar rotational axis remains ``vertical'' but solar North periodically alternates between up and down. 
This has severe consequences on the C2 images as the roll axis imposed by SOHO does not coincide with the C2 reference axis, which points to the center of the Sun: i) a periodic shift of the pixel coordinates of this center, and ii) a change of the stray-light pattern, further complicated by the intrinsic temporal evolution of the instrument performance.
On 29 October 2010 and still on-going, the attitude of SOHO was changed to simplify operation: the reference orientation is fixed to the perpendicular to the ecliptic plane causing the projected direction of the solar rotational axis to oscillate between $\pm$7 \deg 15' around the ``vertical'' direction on the LASCO images. 
The temporal evolution of the roll angle of SOHO is presented in Figure~\ref{fig:roll}. 

The cadence of the ``F+SL'' images is by construction similar to that of the polarization sequences and is given in Figure~3 of \cite{Lamy2020}, which displays the monthly averaged daily rate, one per day until 2008 inclusive and approximately four per day thereafter.

\subsection{Constraints Associated with the Observation of the F-corona from 1\,AU}
     \label{Sec:Const_F} 

\begin{figure}[htpb!]
\begin{center}
\includegraphics[width=\textwidth]{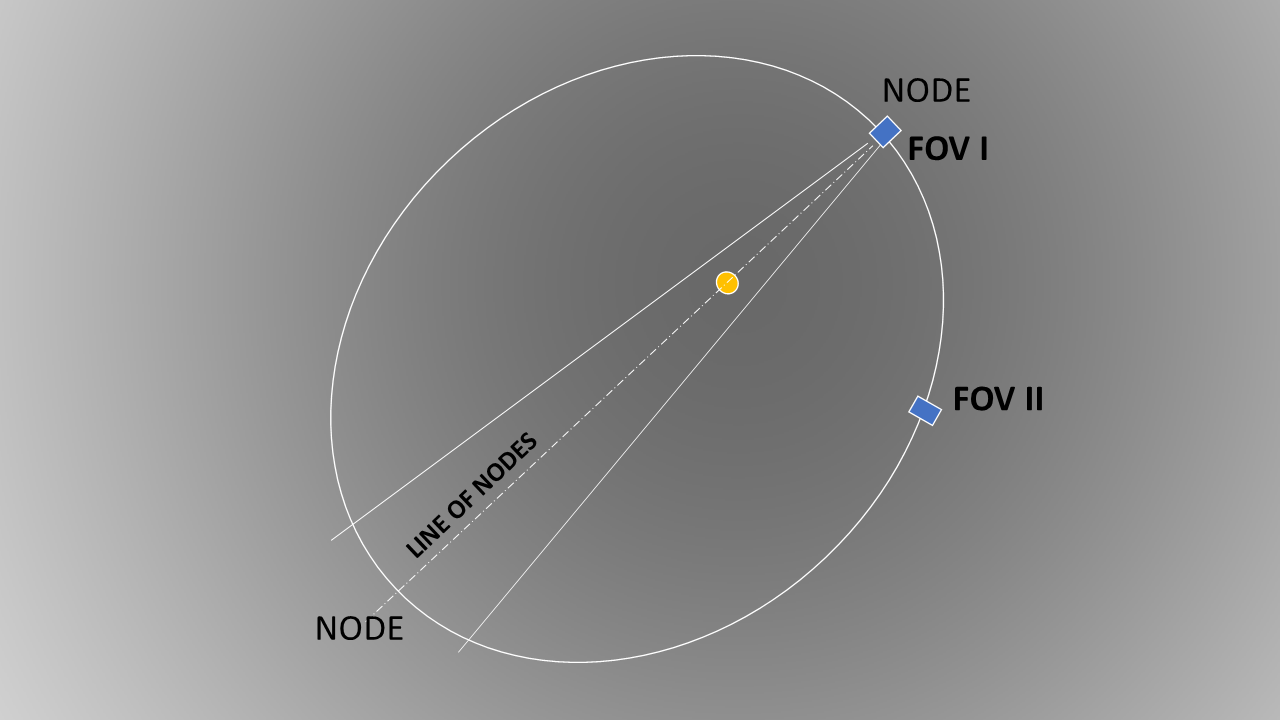}
\includegraphics[width=\textwidth]{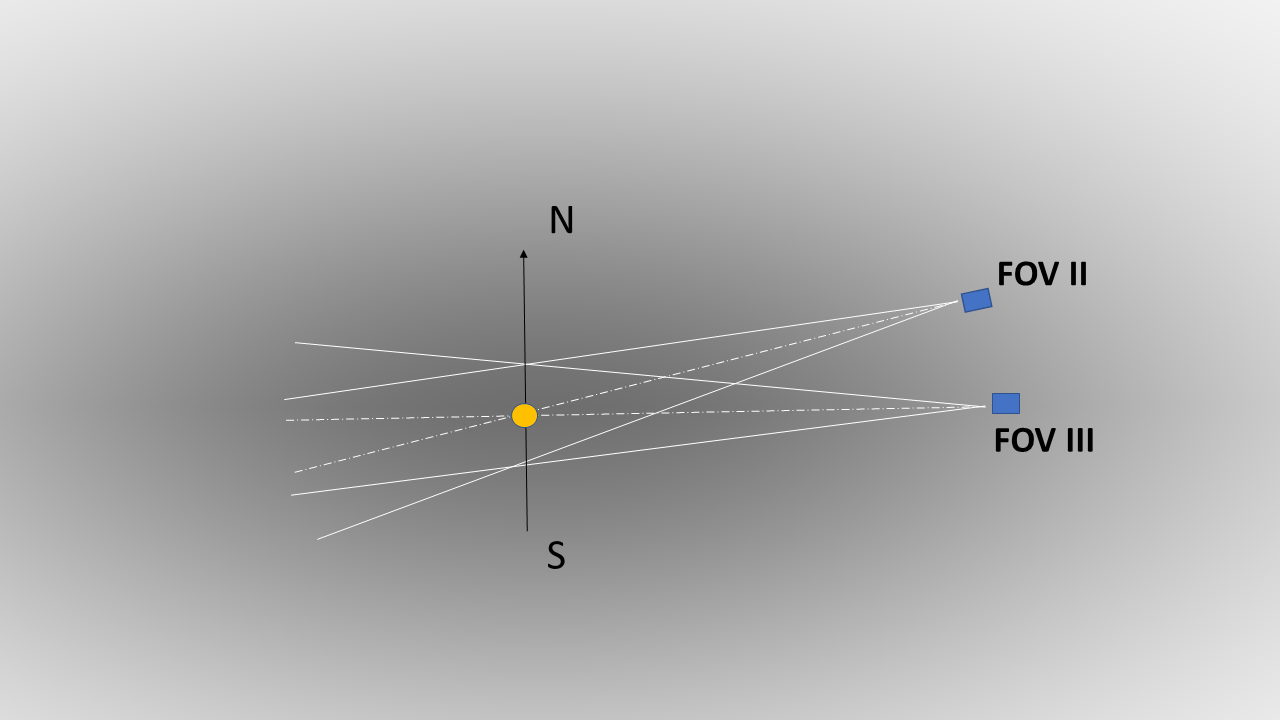}
\caption{Schematic views of the geometry of LASCO observations of the F-corona from different vantage points immersed in the zodiacal cloud represented by the gray background. 
The upper panel shows the orbit of SOHO (white ellipse) in the zodiacal cloud as seen from above its plane of symmetry (PSZC).
The line of nodes defined by the intersection of the orbital plane of SOHO (the Ecliptic plane) and the PSZC is nearly aligned with the major axis of the elliptic orbit.
Its eccentricity is grossly amplified to highlight the difference of heliocentric distances at the two nodes, $\approx$0.973 and $\approx$1.009\,AU. 
FOV I corresponds to an observation at one node whereas FOV II corresponds to an observation in quadrature when SOHO is near its extreme elevation from the plane of symmetry.
This is best seen in the lower panel where this plane is seen edge-on.
FOV III corresponds to FOV I of the upper panel, but artificially rotated by 90$\degr$ to best highlight the different regions of the zodiacal cloud seen from the nodes and from the points in quadrature (\ie along the minor axis of the elliptic orbit).}
\label{fig:orbit3D}
\end{center}
\end{figure}

Unlike the K-corona, which results from scattering of photospheric light by free electrons within a few ten or so solar radii, the F-corona predominantly results from diffraction by dust particles all along the line-of-sight, from the observer to the circumsolar region and even beyond \citep{VdH1947}.
As SOHO travels on its orbit which lies inside the zodiacal cloud, LASCO sees the F-corona from different vantage points with corresponding lines-of-sight sampling different regions of the cloud.
Two main effects are therefore anticipated on the observed radiance of the F-corona: one resulting from the varying Sun--SOHO distance (3.6\,\%) and the other from the oscillation of SOHO about the plane of symmetry of the inner zodiacal cloud (hereafter abbreviated ``PSZC''), which is close to the solar equatorial plane according to \cite{Leinert1990}, whereas the orbit of SOHO lies in the ecliptic plane.
This latter effect is expected to create an imbalance having its largest amplitude in the north and south directions.
Conversely, the orbit of SOHO crosses the symmetry plane twice per year in late June and late December (at the so-called nodes) at which times the F-corona appears quasi-symmetric.
It is interesting to note that the line of nodes is fortuitously closely aligned with the major axis of the elliptic orbit of SOHO so that the two variations are uncoupled to a large extent: the maximum effect of the varying Sun--SOHO distance takes place very near the nodes whereas the maximum effect of the inclination of the symmetry plane is observed in quadrature. 
These two effects are amplified in Figure~\ref{fig:orbit3D} to help in understanding how they affect the LASCO observations of the F-corona.

\subsection{Problems Associated with the Stray Light}
     \label{Sec:Const_SL} 

\begin{figure}[htpb!]
\begin{center}
\includegraphics[width=0.9\textwidth]{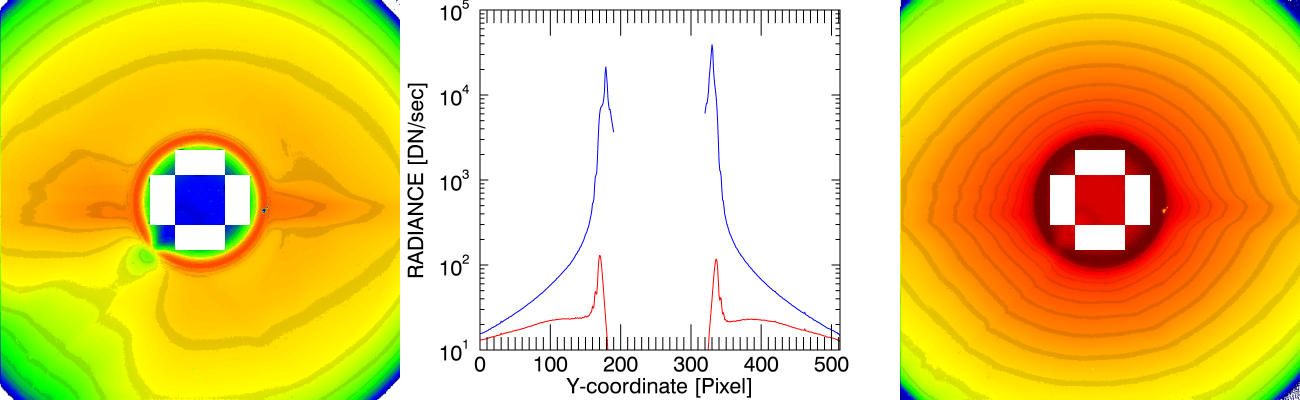}
\caption{Illustration of the vignetting effect on the LASCO-C2 images. 
Left panel: a raw image of 512$\times$512 pixels obtained on 9 January 1997.
Right panel: the same image corrected for the vignetting effect.
Central panel: radial profiles along the south--north direction of the raw image (red curve) and of the corrected image (blue curve).
This illustrates the artificial enhancement of the diffraction fringe.}
\label{fig:fringe}
\end{center}
\end{figure}

The polarimetric analysis of the LASCO-C2 images developed by \cite{Lamy2020} aimed at recovering the images of the polarized radiance $pB$ of the K-corona, and in turn of the electron density. 
Henceforth, the processing included a correction for the instrumental vignetting so that the left-over unpolarized ``F+SL'' images of interest to the present work were corrected for this effect.
This is obviously fine for the F-corona but not for the stray light, which is an internal instrumental effect seen by the unvignetted optical pupil of the coronagraph.
Consequently, the stray light and particularly the diffraction fringe surrounding the occulter are artificially enhanced, thus appearing brighter than in reality (Figure~\ref{fig:fringe}).
This constitutes a serious difficulty that the analysis aimed at restoring the F- and SL-components must cope with, further complicated by the fact that both vary with the Sun--SOHO distance differently, as we shall see.

\section{Procedure of the Restoration of the F-corona}
 \label{Sec:PROC}

Our procedure of restoring the F-corona is quite complex and is composed of three main stages, each involving different steps.
We first present an overview to help in understanding the flow of operations.

\subsection{Overview of the Procedure}
 \label{Sec:OVER} 

\begin{figure}[htpb!]
\begin{center}
\includegraphics[width=\textwidth]{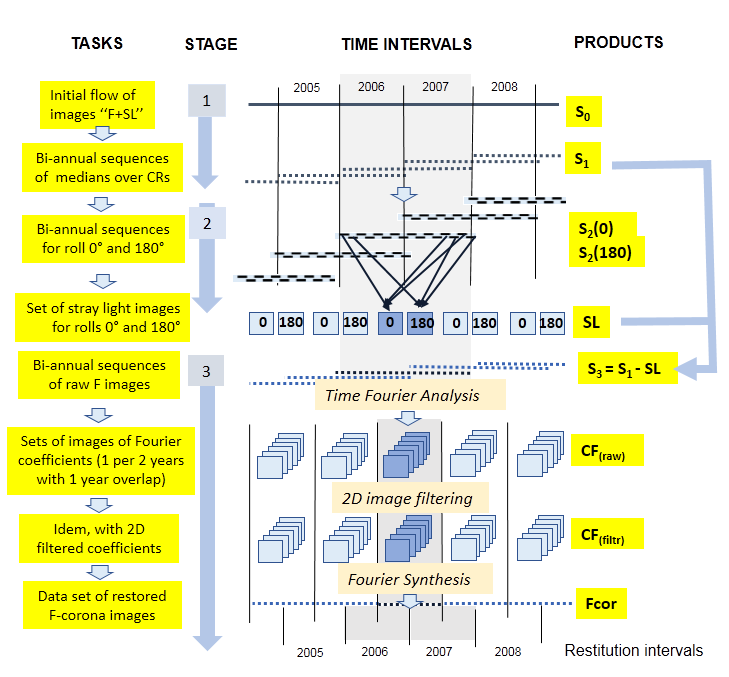}
\caption{Schematic representation of the procedure aimed at separating the F- and SL-components from the ``F+SL'' images produced by the polarimetric analysis (${S_{0}}$ time series).
The upper part of the diagram corresponds to Stages 1 and 2 of the procedure aimed at recovering the stray-light images.
The ${S_{1}}$ time series is obtained by taking the median image for each Carrington rotation.
The ${S_{2}}$ time series results from the application of a two-year sliding window with a time step of one year split into two time series ${S_{2}(0)}$ and ${S_{2}(180)}$ according to the roll angle of SOHO; they are used to generate the images of the stray light.
The lower part of the diagram corresponds to Stage 3 of the procedure aimed at the restoration of the F-corona.
After subtracting the stray light, the ${S_{2}(0)}$ and ${S_{2}(180)}$ are combined in a common coronal reference time series ${S_{3}}$ and its Fourier analysis yields a set of images of the Fourier coefficients ${C_F(raw)}$.
These images are filtered and the resulting coefficients ${C_F(filter)}$ are used to construct the final daily images ``Fcor'' of the F-corona.}
\label{fig:Procedure}
\end{center}
\end{figure}

The analysis of the whole set of the LASCO-C2 polarization sequences by \cite{Lamy2020} performed the separation of the polarized K-corona and the unpolarized component ``F+SL'' composed of the F-corona and the instrumental stray light ``SL''. 
Our procedure of restoring the F-corona makes use of the time series ${S_{0}}$ of these ``F+SL'' images, which extends over 24 years, from 1996 to 2019 inclusive.
As emphasized by \cite{Lamy2020}, the polarimetric technique implemented in the LASCO instrument and based on three identical linear polarizers with orientations at +60$^{\circ}$, 0$^{\circ}$, and -60$^{\circ}$ is far from ideal and the separation is consequently not perfect.
Careful inspection of the ``F+SL'' images indeed revealed the presence of traces or remnants of the K-corona, especially from its brightest structures.
This constitutes a supplementary difficulty that the procedure must cope with.
We present below the three main stages of the procedure schematically summarized in Figure~\ref{fig:Procedure}.

In Stage 1, we perform a global analysis of the ${S_{0}}$ time series aimed at separately determining the variation of the F- and SL-components with the Sun--SOHO distance  and scaling the images accordingly.
We then generate a new shorter time series ${S_{1}}$ by taking the median image for each Carrington rotation, thus attenuating the contamination by K-corona remnants.\\
The main purpose of Stage 2 is the restoration of the stray light.
In a first step, we construct mean images by applying a two-year sliding window with a time step of one year to the ${S_{1}}$ time series to force the F-corona to become artificially symmetric so as to facilitate the F--SL separation.
This results in a new time series ${S_{2}}$ which is split into two time series ${S_{2}(0)}$ and ${S_{2}(180)}$ after 10 July 2003 according to the roll angle of SOHO.  
In a second step, we construct a parametric model of this symmetric F-corona.
The third step is devoted to the final restoration of images of the stray light after defining time intervals during which they may be considered constant.\\
In Stage 3, we perform the restoration of the F-corona starting from a new time series ${S_{3}}$ which combines the ${S_{2}(0)}$ and ${S_{2}(180)}$ time series.
A Fourier analysis in the temporal domain for each point of the corona $(i,j)$  yields a set of images of the Fourier coefficients ${C_F(raw)}$, which are subsequently filtered in the spatial domain. 
The resulting coefficients ${C_F(filter)}$ are finally used to construct the final daily images of the F-corona, named ``Fcor'', throughout the first 24 years of the mission [1996\,--\,2019].\\

Our procedure is reminiscent of that implemented by \cite{Stenborg2017} for estimating (and removing) the background in STEREO HI-1 images in the sense that both rely on the different morphology and spatial frequencies of the components to be separated.
However, we point out that their application to a LASCO-C2 image (their Figure~14) has nothing to do with this present work.
Their ``background'', which most likely includes the F-corona and the background (smooth) K-corona is used to construct the ratio with the original image.
In essence, this is equivalent to an unsharp-masking operation aimed at revealing the K-corona structures and a CME at the expense of losing the photometry.

\subsection{Stage 1: Global Analysis of the  Time Series of ``F+SL'' Images}
\label{Sec:Stage1}

As already mentioned in Section~\ref{Sec:Const}, the temporal variation of the Sun--SOHO distance $d$ affects the radiance of both the F-corona and the stray light.
The F-corona increases as $d$ decreases and we expect a power law variation $d^{-\scalebox{1.}{$\nu$}_{\scriptscriptstyle\rm F}}$ close to that found by the two {\it Helios} space probes $d^{-2.3 \pm 0.05}$ from measurements between 0.3 and 1\,AU \citep{Leinert1981}.
The stray light increases as well due to the change in the apparent size of the Sun, and we will see later that its variation $d^{-\scalebox{1.}{$\nu$}_{\scriptscriptstyle\rm SL}}$ is much steeper than that of the F-corona since it is prominently controlled by small-angle diffraction by the occulter.

\begin{figure}[htpb!]
\begin{center}
\includegraphics[width=0.8\textwidth]{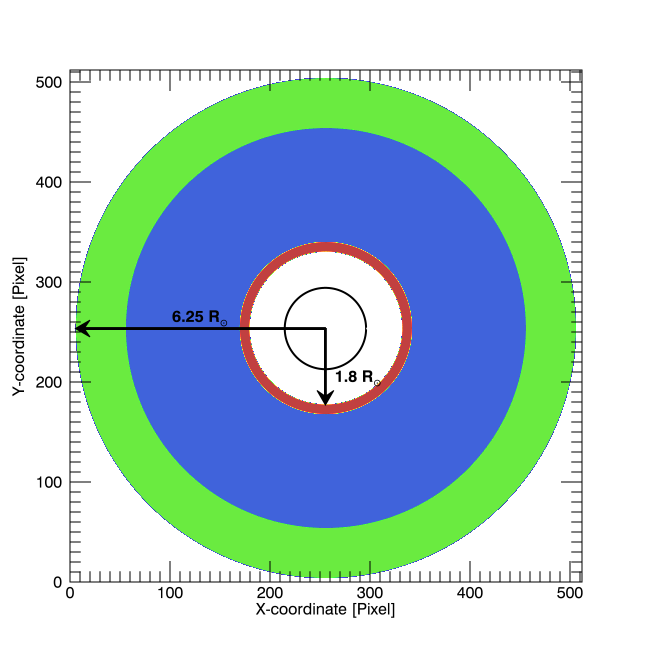}
\caption{Illustration of the three concentric rings applied to the ``F+SL'' images of 512$\times$512  pixels to study the photometric variation of the F-corona and the stray light with the Sun--SOHO distance.
The black circle represents the solar disk.
The red ring isolates the diffraction fringe.
In the blue ring, the F-corona is affected by stray effects whereas in the green ring, these effects are presumed to be absent.}
\label{fig:3rings}
\end{center}
\end{figure}

In order to build an homogeneous time series of ``F+SL'' images, we must first determine these variations, that is the two power exponents $\nu_{\scriptscriptstyle\rm F}$ and $\nu_{\scriptscriptstyle\rm SL}$, and normalize the images according to which component F or SL will be studied; this is the aim of the first step of Stage 1.   
To do so, we define three circular regions or rings as illustrated in Figure~\ref{fig:3rings}.
\begin{itemize}
\item A narrow ring isolating the diffraction fringe defined by its inner and outer radii of 76.5 and 86 pixels, respectively.
\item A broad intermediate ring extending radially from 86 to 200 pixels where the F-corona may still be affected by remnants of stray light and K-corona.
\item A narrow outer ring extending radially from 200 to 250 pixels where the F-corona is presumed to be nearly free of any stray effects since the contributions from the stray light and the K-corona are expected to be vanishingly small.
\end{itemize}

\begin{figure*}
\vspace{0.5cm}
\noindent
\centering
\includegraphics[height=\textwidth, width=0.91\textheight,angle=90]{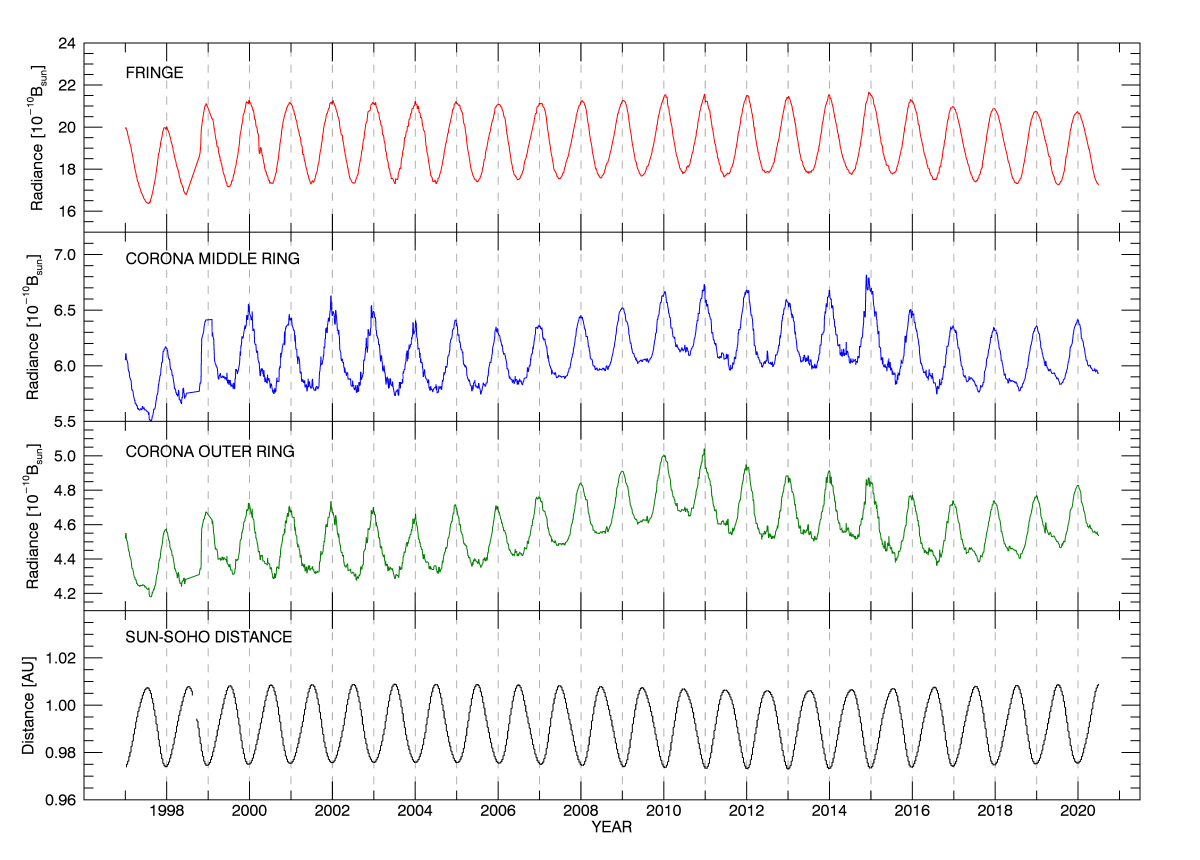}
\caption{Temporal variation of the radiances of the ``F+SL'' images integrated in the three rings defined in Figure~\ref{fig:3rings} over 24 years (upper three panels).
The colors correspond to the rings depicted in Figure~\ref{fig:3rings}.
The lower panel displays the variation of the Sun--SOHO distance.}
\label{fig:FluxesAll}
\end{figure*}

Integrating the radiances in these rings has the benefit of removing the periodic North--South asymmetry affecting the F-corona when SOHO moves back and forth about the symmetry plane of the zodiacal cloud, so that their temporal variations prominently reflect the effect of the varying Sun--SOHO distance.
The integrations in the intermediate and outer rings are performed on the ``F+SL'' images, since they properly record the F-corona, whereas that in the inner ring is performed on the vignetted ``F+SL'' images, since they correctly record the diffraction fringe as explained in Section~\ref{Sec:Const_SL}.
The temporal variations of the three integrated radiances are displayed in Figure~\ref{fig:FluxesAll}, and their periodic oscillations are clearly in phase with the yearly variations of the Sun--SOHO distance. 
The long-term trend of the radiance of the fringe is remarkably stable with only slight variations.
After the recovery of SOHO in 1998, it increased by 5\,\%, probably a consequence of a slight offset between the pre- and post-pointings.
Thereafter, the radiance remained at a quasi-constant level until 2010 when the general level increased by 1.2\,\% and remained constant until 2015.
The next three years witnessed a slight continuous decrease which leveled off during the following two years.
Compared with the [1999\,--\,2010] time interval, the present general level is reduced by 2.4\,\%.
The long-term trends of the radiance in the middle and outer rings follow similar evolutions: stable until 2006, then a progressive increase culminating in 2011 with a 5\,\% increase, followed by a decrease in 2016 and a turnover in 2019, all these fluctuations remaining at the level of a few percent.
The influence of the K-corona residuals mentioned in Section~\ref{Sec:OVER} may be perceived by the irregularities in the amplitudes of the yearly variations during the two solar maxima and the simultaneous presence of high-frequency fluctuations best seen in the minima of the curves.
In addition, the prominent peak present in the case of the middle ring at the end of 2014 reflects the anomalous surge that affected the K-corona as described by \cite{Lamy2017}.

\begin{figure}[htpb!]
\begin{center}
\includegraphics[width=0.9\textwidth]{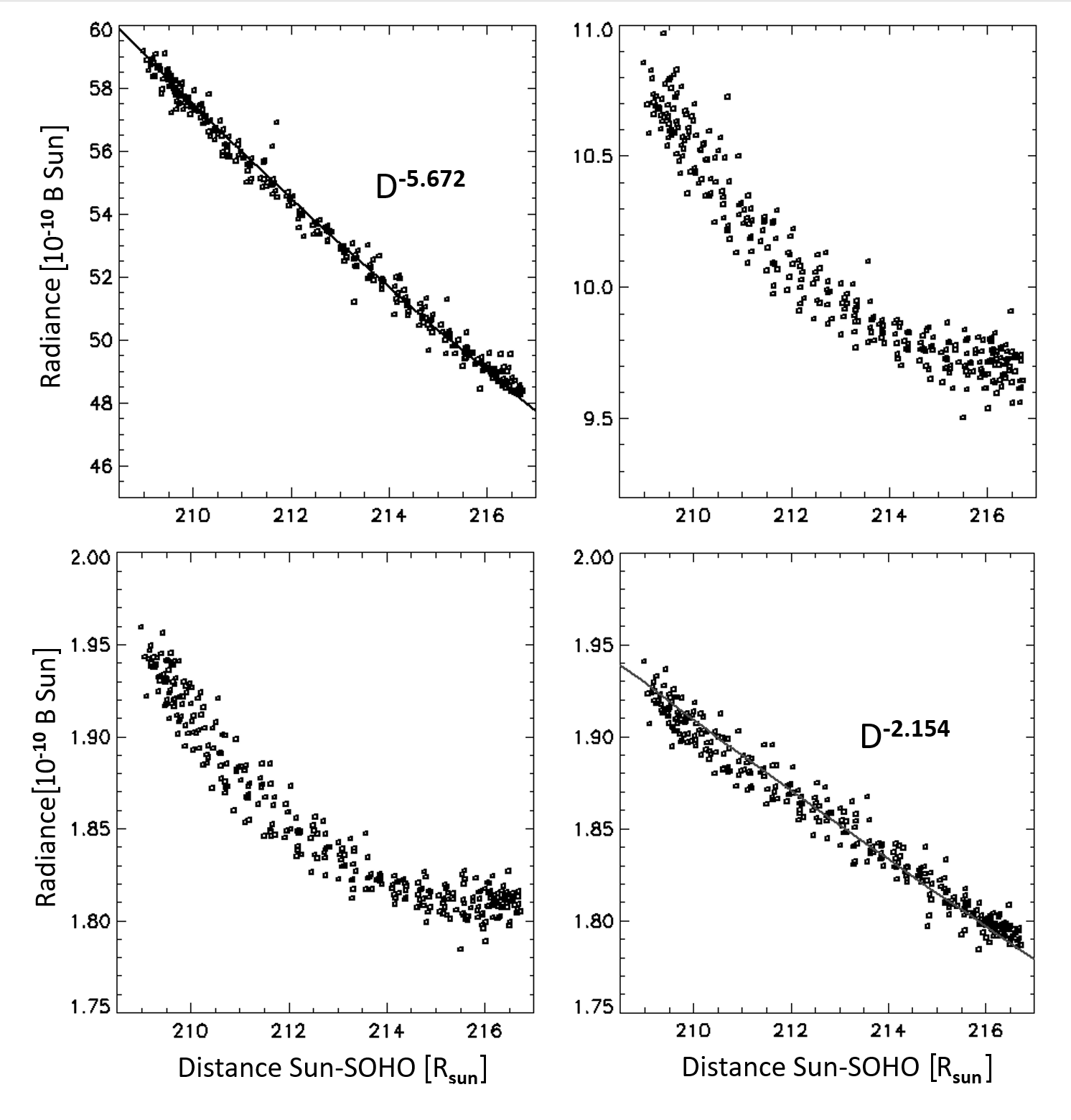}
\caption{Scatter plots of the integrated radiance in the three rings versus the Sun--SOHO distance on a log--log scale: inner fringe ring (upper-left panel), middle ring (upper- right panel), and outer ring (lower-left panel).
The lower-right panel displays the effect of de-trending the radiance in the outer ring by removing the semi-annual variation.}
\label{fig:scatterplot}
\end{center}
\end{figure}

Figure~\ref{fig:scatterplot} displays the three scatter plots corresponding to the three rings where the logarithm of the integrated radiance is plotted versus the logarithm of the Sun--SOHO distance.
Whereas the fringe exhibits a clear linear trend allowing us to unambiguously derive a power exponent $\nu_{\scriptscriptstyle\rm SL}$ = 5.672, the behaviour of the two others is quite complex and an in-depth analysis revealed that it results from the superposition of the long-term temporal evolution and of the combination of the annual (SOHO's orbit) and semi-annual (node crossings) variations. 
The second effect basically replaces the expected linear curve by a ``Lissajous type'' curve, whereas the first effect blurs the scatter plot.
Disentangling these effects required an extra analysis starting with the determination and the subtraction of the long-term trends.
This was achieved by either tracking the extrema of the temporal variations or applying a running average with a window of one year; the two approaches perform equally well.
Then the separation of the annual and semi-annual variations was achieved by iteratively fitting a power-law of the Sun--SOHO distance to the temporal variations.
This quickly converged and resulted in power-law variations as illustrated by the case of the outer ring displayed in the lower-right panel of Figure~\ref{fig:scatterplot}.
This procedure yielded robust determinations of the power exponents: 2.154 for the outer ring and 2.926 for the middle one. 

\begin{figure}[htpb!]
\begin{center}
\includegraphics[width=0.9\textwidth]{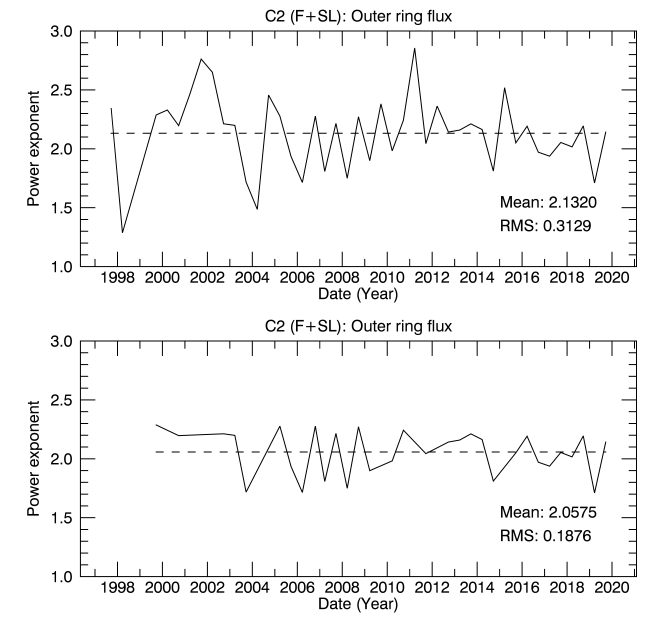}
\caption{Temporal evolution of the power exponent $\nu_{\scriptscriptstyle\rm F}$ of the variation of the radiance of the ``F+SL'' images integrated in the outer ring.
The upper panel displays the complete data set and the lower one excludes the largest fluctuations, see text for detail.}
\label{fig:nu_semiannual}
\end{center}
\end{figure}

An alternative procedure was applied consisting in considering only the radiance values at the nodes, exploiting the fact that they take place at nearly the extreme values of the Sun--SOHO distance as pointed out in Section~\ref{Sec:Const}.
The nodes further offer an advantage as C2 sees the  \emph{same} volume of the zodiacal cloud from two opposite vantage points so that the variation of the radiance between  consecutive nodes can be safely attributed to the varying Sun--SOHO distance.
This approach bypasses the tasks of determining the long-term trend and of disentangling the annual and semi-annual variations.
Individual power exponent values were thus derived every half-year and averaging them produced global values over 24 years.
Figure~\ref{fig:nu_semiannual} illustrates this determination for the outer ring yielding a power-law exponent of 2.132 to be favorably compared with 2.154 from the first method.
It is interesting to note that, except for the first years when the accidental interruptions of SOHO may have introduced a bias, the largest fluctuations are recorded during the maxima of SC 23 and 24, possibly suggesting some influence of the K-corona at those times even in the outer ring.
Removing these spurious data points reduces the power exponent to 2.058, which is probably a more realistic value for the F-corona. 
The standard deviation appears quite large, 0.31 in the nominal case and 0.19 after removing the ``outliers'', underlying the sensitivity of the power exponent to the data.

Let us come back to the large difference between the values of the power exponent found for the middle and outer rings.
This was investigated by subdividing the middle ring into four sub-rings of approximately the same width, and we did find a progressive reduction of the exponent from 3.685 in the innermost sub-ring to the value mentioned above in the outer ring.
There is little doubt that we are witnessing the diminishing influence of the stray light with increasing solar elongation, starting from an exponent of 5.672 for the fringe itself.
This complicated the adoption of a single power-law variation for the F-corona in the ``F+SL'' images, and we made a compromise by retaining an average value $\nu_{\scriptscriptstyle\rm F}$ = 2.3 as obtained in the sub-ring located half-way in the C2 \fovnospace.

The final step of Stage 1 consisted in correcting the ``F+SL'' images for the varying Sun--SOHO distance (according to the function $d^{-\scalebox{1.}{$\nu$}_{\scriptscriptstyle\rm F}}$) and reducing their large number by taking the median images calculated for each Carrington rotation.
This procedure produced a time series of 312 images called ${S_1}$ and offered the additional benefit of mitigating the effects of the K-corona residuals. 

\subsection{Stage 2: restoration of the Stray Light}
\label{Sec:Stage2}

Due to the stability of both the pointing of SOHO and the C2 instrument, we expect the stray light to be stable certainly not over the whole mission, but on time-scales much larger than a Carrington rotation so that the sampling of the ${S_1}$ time series is not relevant and necessary for its restoration.
Another aspect comes into play as the F-corona must be regularized or smoothed as much as possible to facilitate the F/SL separation.

The first step of Stage 2 consisted in generating a new reduced time series ${S_2}$ satisfying the above prescriptions, and this was achieved by constructing mean images using a two-year sliding window with a time step of one year.
Each image denoted $I_M (i,j)$ with $i,j =\{1\dots 512\}$ is therefore the mean of 26 images of the ${S_1}$ time series.
This process removes the yearly variation of the F-corona and creates a symmetric ``mean F-corona'' superimposed on a mean stray light. 
This ``mean F-corona'' has no physical reality but very much helps for the F/SL separation. 

Starting on 10 July 2003, the quarterly SOHO rolls imposed the division of  ${S_{2}}$ into two separate sub-series ${S_2(0)}$ and ${S_2(180)}$, which were processed independently. 
Indeed, preliminary investigations revealed a systematic change of the location of the center of the Sun between the two roll cases (the SOHO roll axis is evidently not strictly aligned with the C2 pointing axis) and slight changes of the stray-light pattern.
In order to ensure a good continuity between successive years, the pre-roll 26 images were recovered by implementing the two-year window with an overlap of one year, which smooths the transition.

At this point, we had to consider the question of the stability of the stray light, that is to define time intervals during which it can be considered constant, and separately for the two roll cases.
This was achieved by visual inspection of running differences between successive images of the ${S_2(0)}$ and ${S_2(180)}$ time series. 
Therefore, at the end of this first step, we obtained: 
i) a set of images $I_M (i,j)$  where a symmetric ``mean F-corona'' and an enhanced (by the vignetting correction) stray light are superimposed, and 
ii) a set of time intervals during which the stray light may be considered constant (Table~\ref{tab:dates_validity}).
Note that any K-corona residuals are expected to be further reduced as a consequence of the double averaging resulting from the constructions of the ${S_{1}}$ and  ${S_{2}}$ time series.

\begin{longtable}{ccc}
\caption{Start times in both Modified Julian Date (MJD) and calendar date of the intervals of validity of the different stray-light images for the years 1996 to 2019.}\\
\noalign{\smallskip}\hline\noalign{\smallskip}
MJD       & Calendar date    & File name \\
\noalign{\smallskip}\hline\noalign{\smallskip}
50126.588 & 1996-02-13 14:06 & SL\_1996\_1998\_000 \\
51099.000 & 1998-10-13 00:00 & SL\_1999\_2000\_000 \\
51879.003 & 2000-12-01 00:00 & SL\_2001\_2002\_000 \\
52312.764 & 2002-02-07 18:20 & SL\_2002\_2003\_000 \\
52828.542 & 2003-07-08 13:00 & SL\_2003\_2004\_180 \\
52919.465 & 2003-10-07 11:10 & SL\_2003\_2004\_000 \\
53003.817 & 2003-12-30 19:36 & SL\_2003\_2004\_180 \\
53094.775 & 2004-03-30 18:36 & SL\_2003\_2004\_000 \\
53180.612 & 2004-06-24 14:41 & SL\_2004\_2005\_180 \\
53270.956 & 2004-09-22 22:56 & SL\_2004\_2005\_000 \\
53362.891 & 2004-12-23 21:23 & SL\_2004\_2005\_180 \\
53452.879 & 2005-03-23 21:05 & SL\_2004\_2005\_000 \\
53539.762 & 2005-06-18 18:17 & SL\_2005\_2006\_180 \\
53622.515 & 2005-09-09 12:21 & SL\_2005\_2006\_000 \\
53720.937 & 2005-12-16 22:29 & SL\_2005\_2006\_180 \\
53808.171 & 2006-03-14 04:06 & SL\_2005\_2006\_000 \\
53893.673 & 2006-06-07 16:09 & SL\_2006\_2007\_180 \\
53989.743 & 2006-09-11 17:49 & SL\_2006\_2007\_000 \\
54075.842 & 2006-12-06 20:12 & SL\_2006\_2007\_180 \\
54164.996 & 2007-03-05 23:54 & SL\_2006\_2007\_000 \\
54251.062 & 2007-05-31 01:29 & SL\_2007\_2008\_180 \\
54344.075 & 2007-09-01 01:48 & SL\_2007\_2008\_000 \\
54425.117 & 2007-11-21 02:48 & SL\_2007\_2008\_180 \\
54519.884 & 2008-02-23 21:12 & SL\_2007\_2008\_000 \\
54607.032 & 2008-05-21 00:46 & SL\_2008\_2009\_180 \\
54700.695 & 2008-08-22 16:40 & SL\_2008\_2009\_000 \\
54790.397 & 2008-11-20 09:31 & SL\_2008\_2009\_180 \\
54877.754 & 2009-02-15 18:05 & SL\_2008\_2009\_000 \\
54963.886 & 2009-05-12 21:15 & SL\_2009\_2010\_180 \\
55056.886 & 2009-08-13 21:15 & SL\_2009\_2010\_000 \\
55144.618 & 2009-11-09 14:49 & SL\_2009\_2010\_180 \\
55231.879 & 2010-02-04 21:05 & SL\_2009\_2010\_000 \\
55316.942 & 2010-04-30 22:36 & SL\_2010\_2011\_180 \\
55410.811 & 2010-08-02 19:27 & SL\_2010\_2011\_000 \\
55498.939 & 2010-10-29 22:32 & SL\_2010\_2011\_180 \\
55586.642 & 2011-01-25 15:24 & SL\_2010\_2011\_000 \\
55673.755 & 2011-04-22 18:07 & SL\_2011\_2012\_180 \\
55764.833 & 2011-07-22 19:59 & SL\_2011\_2012\_000 \\
55854.971 & 2011-10-20 23:18 & SL\_2011\_2012\_180 \\
55944.300 & 2012-01-18 07:12 & SL\_2011\_2012\_000 \\
56028.940 & 2012-04-11 22:33 & SL\_2011\_2012\_180 \\
56032.342 & 2012-04-15 08:12 & SL\_2012\_2013\_180 \\
56120.947 & 2012-07-12 22:43 & SL\_2012\_2013\_000 \\
56210.018 & 2012-10-10 00:25 & SL\_2012\_2013\_180 \\
56299.841 & 2013-01-07 20:11 & SL\_2012\_2013\_000 \\
56400.684 & 2013-04-18 16:24 & SL\_2012\_2013\_180 \\
56410.084 & 2013-04-28 02:00 & SL\_2013\_2014\_180 \\
56476.742 & 2013-07-03 17:48 & SL\_2013\_2014\_000 \\
56565.917 & 2013-09-30 22:00 & SL\_2013\_2014\_180 \\
56663.919 & 2014-01-06 22:03 & SL\_2013\_2014\_000 \\
56770.659 & 2014-04-23 15:48 & SL\_2013\_2014\_180 \\
56831.882 & 2014-06-23 21:10 & SL\_2014\_2015\_000 \\
56922.910 & 2014-09-22 21:50 & SL\_2014\_2015\_180 \\
57015.709 & 2014-12-24 17:00 & SL\_2014\_2015\_000 \\
57164.179 & 2015-05-22 04:17 & SL\_2014\_2015\_180 \\
57185.855 & 2015-06-12 20:31 & SL\_2015\_2016\_000 \\
57276.992 & 2015-09-11 23:48 & SL\_2015\_2016\_180 \\
57366.876 & 2015-12-10 21:01 & SL\_2015\_2016\_000 \\
57452.004 & 2016-03-05 00:05 & SL\_2015\_2016\_180 \\
57542.903 & 2016-06-03 21:40 & SL\_2016\_2017\_000 \\
57632.868 & 2016-09-01 20:49 & SL\_2016\_2017\_180 \\
57721.788 & 2016-11-29 18:55 & SL\_2016\_2017\_000 \\
57807.840 & 2017-02-23 20:09 & SL\_2016\_2017\_180 \\
57897.788 & 2017-05-24 18:55 & SL\_2017\_2018\_000 \\
57987.863 & 2017-08-22 20:42 & SL\_2017\_2018\_180 \\
58077.754 & 2017-11-20 18:06 & SL\_2017\_2018\_000 \\
58163.826 & 2018-02-14 19:50 & SL\_2017\_2018\_180 \\
58253.822 & 2018-05-15 19:44 & SL\_2018\_2019\_000 \\
58343.723 & 2018-08-13 17:21 & SL\_2018\_2019\_180 \\
58436.747 & 2018-11-14 17:56 & SL\_2018\_2019\_000 \\
58519.826 & 2019-02-05 19:50 & SL\_2018\_2019\_180 \\
58610.691 & 2019-05-07 16:35 & SL\_2018\_2019\_000 \\
58582.191 & 2019-08-04 16:35 & SL\_2018\_2019\_180 \\
58792.749 & 2019-11-05 17:59 & SL\_2018\_2019\_000 \\
58877.856 & 2020-01-29 20:32 & SL\_2018\_2019\_180 \\
\noalign{\smallskip}\hline\noalign{\smallskip}
\label{tab:dates_validity}
\end{longtable}

The second step of Stage 2 consisted in constructing a model of the ``mean F-corona'' exploiting its properties of symmetry with respect to the x- (east--west) and y- (south--north) axes and the fact that the stray-light pattern is prominently dominated by the narrow diffraction fringe with a faint extension up to $\approx$160 pixels from the center, thus leaving the outer part nearly free of parasitic light. 
The procedure started by taking the logarithm of the radiance of the $I_M (i,j)$ images and transforming them in polar coordinates ($\rho$, $\theta$) whose origin is at the center of symmetry of the ``mean F-corona''.
This center was defined by cross-correlating the direct and reversed radial profiles along the x- and y-axes, excluding the inner zone where the stray light dominates. 
Circular profiles were extracted from each polar transformed images at constant radii $\rho_k$ and fitted with polynomials of degree six retaining only the even exponents:

\begin{eqnarray*}
      P (\theta,\rho_k)  & = & a_0(\rho_k)+a_2(\rho_k)\theta^2+a_4(\rho_k)\theta^4+a_6(\rho_k)\theta^6 .east
\end{eqnarray*}

\begin{figure}[htpb!] 
\begin{center}
\includegraphics[width=0.9\textwidth]{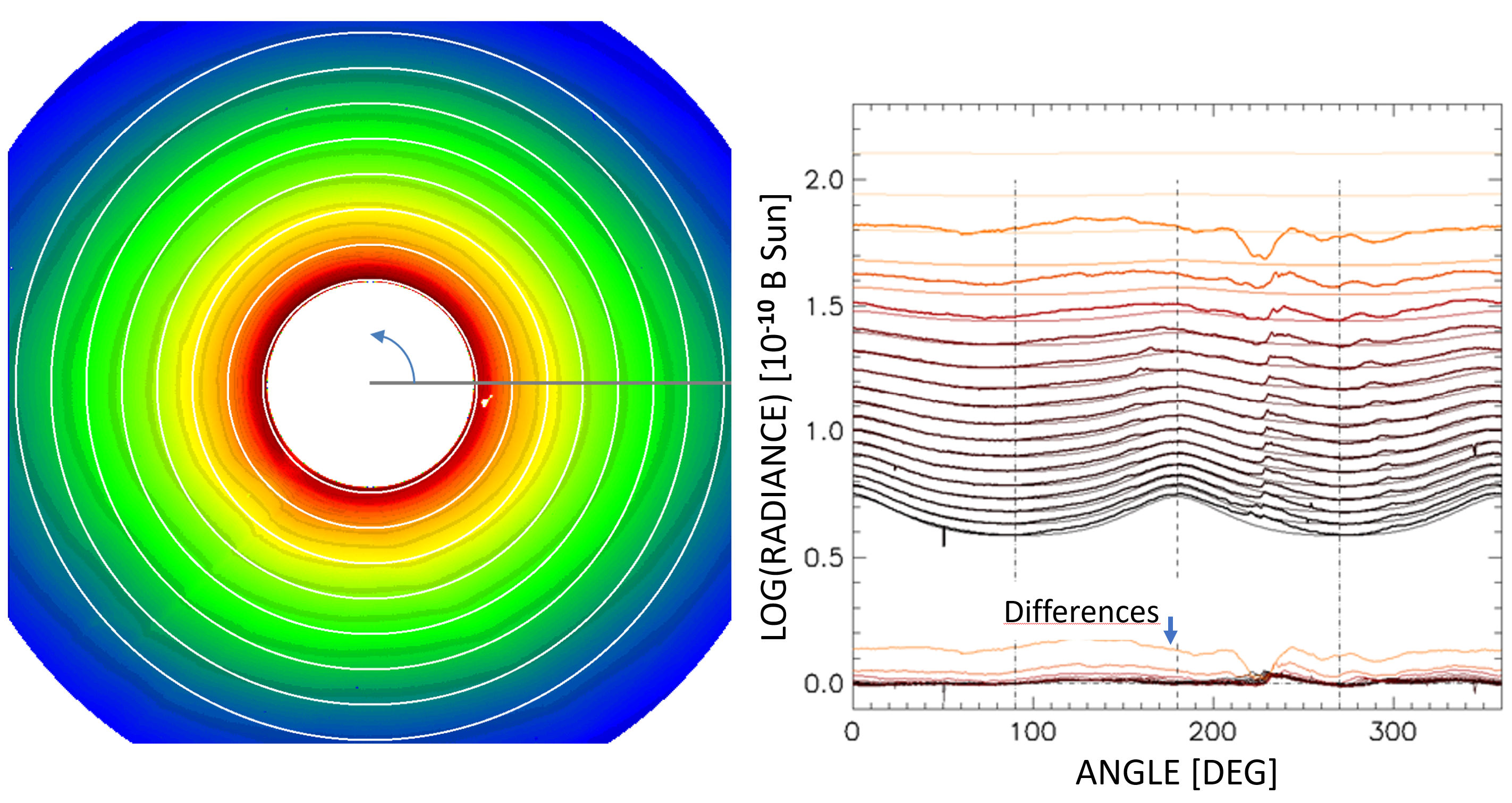}
\caption{Image of a ``mean F-corona'' with a subset of overlaid white circles (left panel) used to extract the polar profiles shown in the right panel.
The polar development starts from the horizontal right radius (black line on the image) and proceeds counterclockwise.
The raw profiles are represented by thick lines and the restored ones using the polynomial fitting procedure by thin lines.
Differences between the logarithm of the two corresponding profiles are displayed below in the 0.0 to 0.3 range.}
\label{fig:FSL_cart_pol}
\end{center}
\end{figure}

The assumed symmetry of the ``mean F-corona'' allowed us restricting the fits to half profiles, from one maximum (east side) to the other (west side), and we favored the northern hemisphere to avoid the sector contaminated by stray effects associated with the pylon.
Additional constraints were introduced in the fitting process, notably the continuity at the maxima (first derivatives equal to zero) and the positivity of the residuals ``raw profiles minus fitted profiles''. 
This latter conditions in fact imposed that the fitted profiles are lower envelopes of the raw profiles (Figure~\ref{fig:FSL_cart_pol}).
The variations of the coefficients $a_{2i}(\rho_k)$ with $i = 0, 1, 2, 3$  as functions of $\rho_k$ were analyzed and all displayed a clear transition from the regime of the F-corona to that of the diffraction fringe, so that the former regime could be extrapolated inward to recover the corrected ``mean F-corona'' down to the inner limit of the \fovnospace.
These extrapolated coefficients were used to construct new polar images $P(\theta,\rho)$, which were then transformed back to Cartesian images yielding the corrected ``mean F-corona'' $F_0 (i, j)$.
A formal and detailed description of this procedure is given by \cite{Llebaria2012}.

The third and final step of Stage 2 consisted in generating the set of stray-light images by subtracting the $F_0 (i, j)$ images from the $I_M (i,j)$ images, and removing the effect of the vignetting correction $\mathcal{V}(i,j)$ as explained in Section~\ref{Sec:Stage1}:
\begin{eqnarray*}
                                              SL (i, j) & = & [I_M (i, j) – F_0 (i, j)] \times \mathcal{V}(i,j) . 
\end{eqnarray*}

\begin{figure}[htpb!] 
\begin{center}
\includegraphics[width=\textwidth]{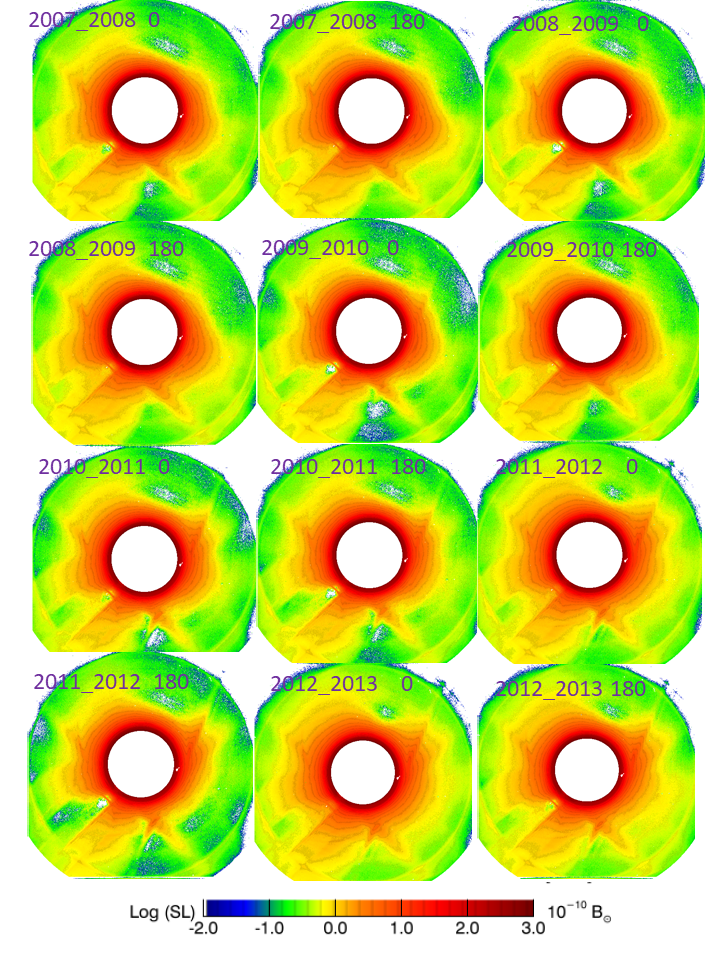}
\caption{Images of the LASCO-C2 stray light over the time interval 1996 to 2007.}
\label{fig:SL1}
\end{center}
\end{figure}

\begin{figure}[htpb!] 
\begin{center}
\includegraphics[width=\textwidth]{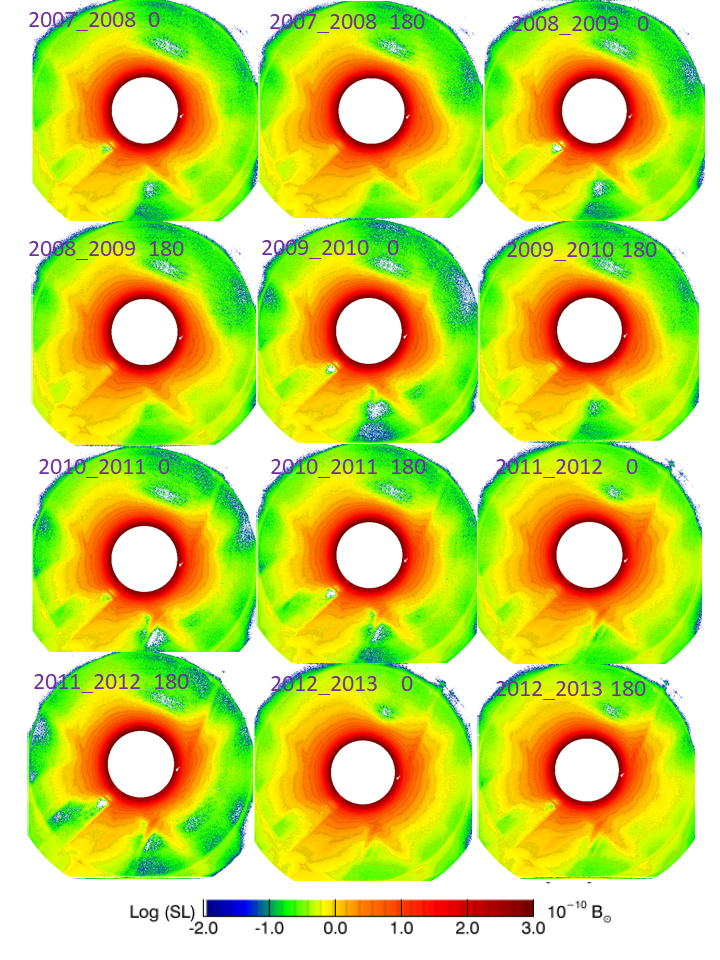}
\caption{Images of the LASCO-C2 stray light over the time interval 2007 to 2013.}
\label{fig:SL2}
\end{center}
\end{figure}

\begin{figure}[htpb!] 
\begin{center}
\includegraphics[width=\textwidth]{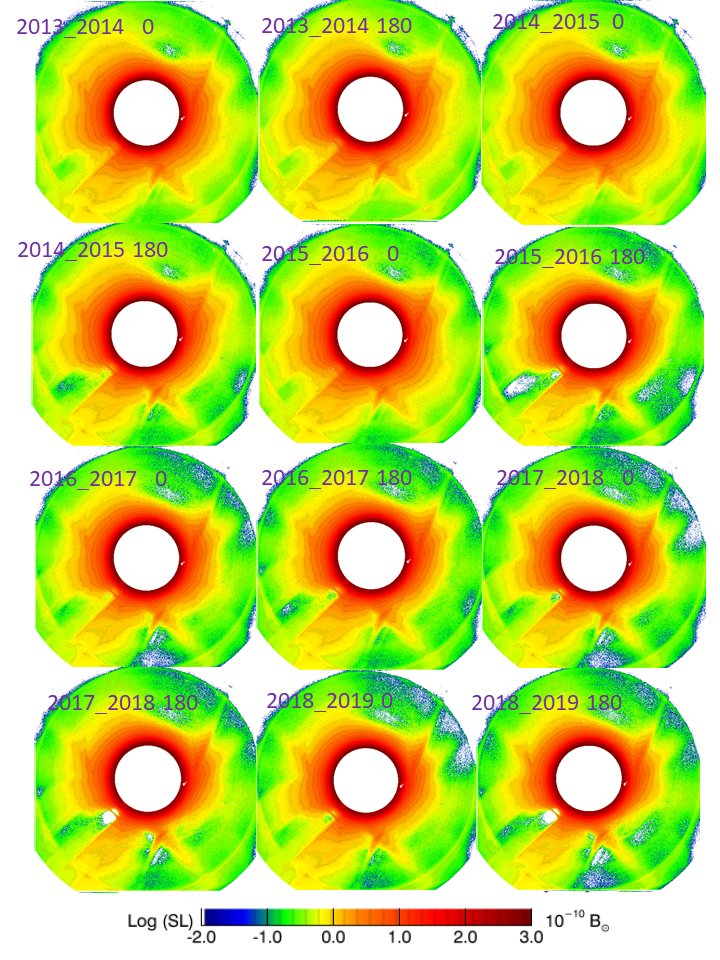}
\caption{Images of the LASCO-C2 stray light over the time interval 2013 to 2019.}
\label{fig:SL3}
\end{center}
\end{figure}

Altogether, we produced a set of 36 stray-light images covering 24 years of LASCO-C2 operation from 1996 to 2019 inclusive. 
They are displayed in Figures~\ref{fig:SL1}, \ref{fig:SL2}, and \ref{fig:SL3}, each one including twelve images.
Table~\ref{tab:dates_validity} lists the dates of change of stray light, in other terms, the time intervals of the validity of each image.
Note the systematic changes at each new roll maneuver.

\begin{figure}[htpb!] 
\begin{center}
\includegraphics[width=\textwidth]{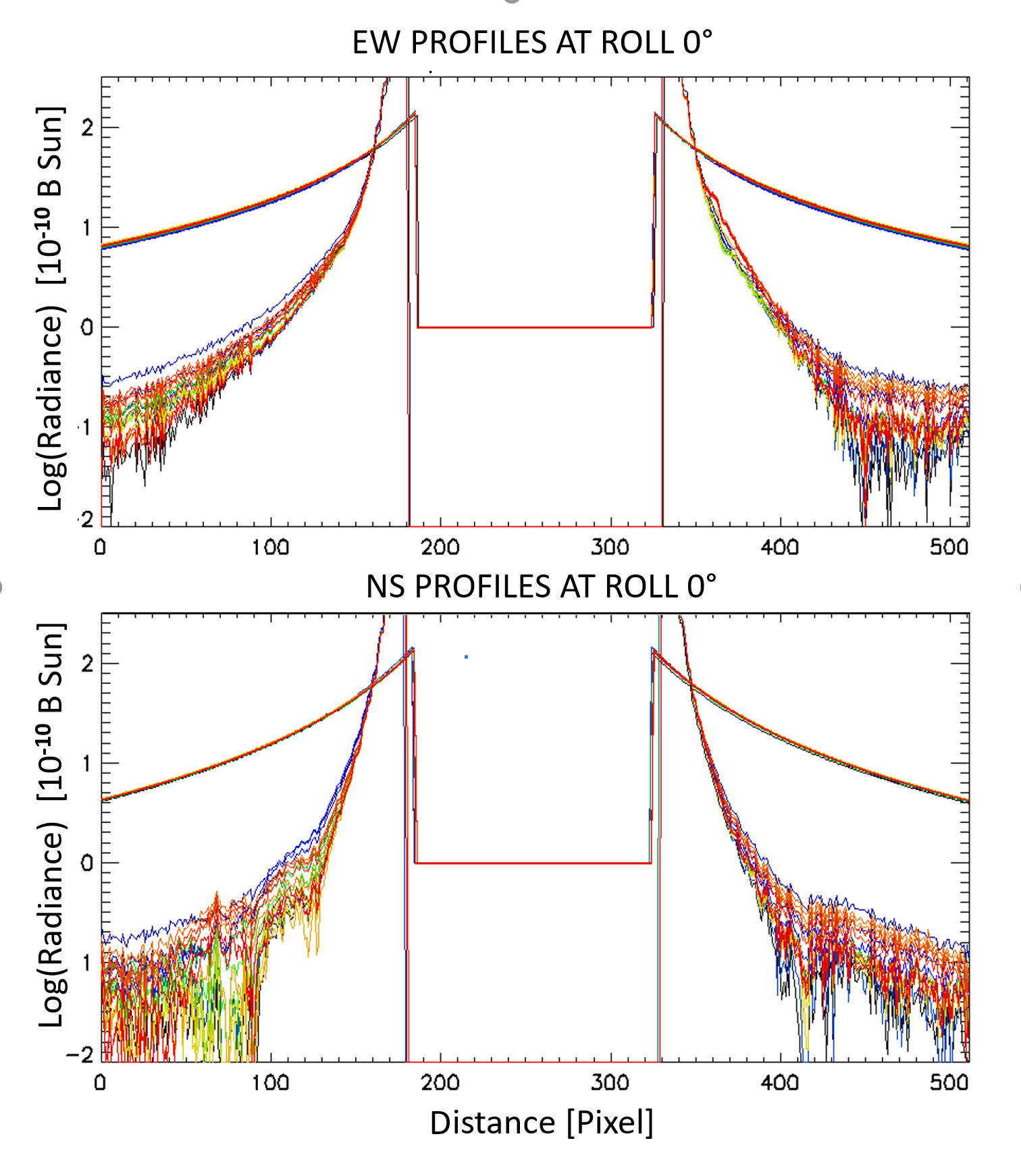}
\caption{East--west (upper panel) and south--north (lower panel) radial profiles of the stray light (lower set of curves) and of the F-corona (upper set of indistinguishable curves) over the [2004\,--\,2018] time interval in the case of the SOHO roll angle of 0$\degr$.}
\label{fig:Bf_SL_000}
\end{center}
\end{figure}

\begin{figure}[htpb!] 
\begin{center}
\includegraphics[width=\textwidth]{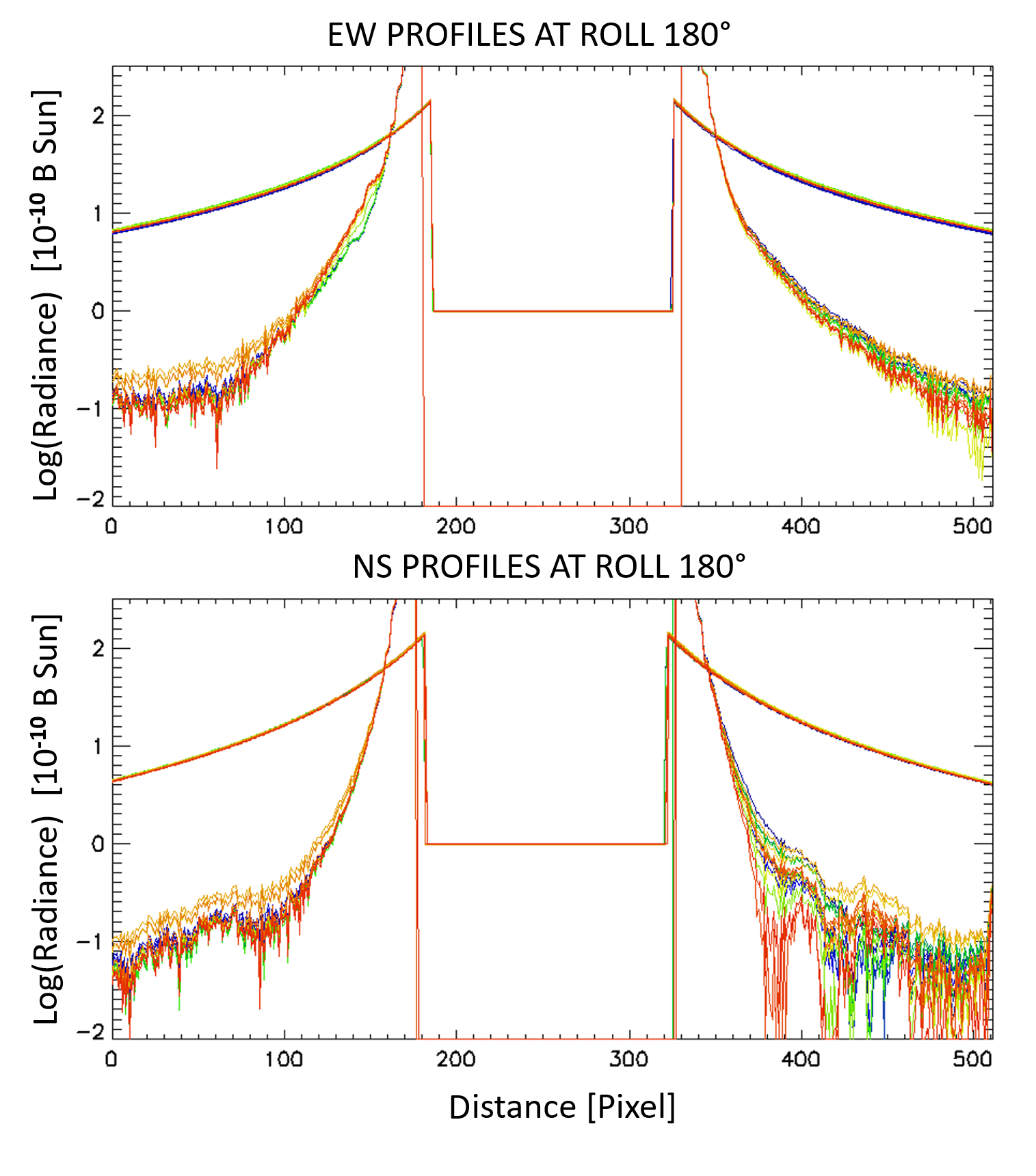}
\caption{East--west (upper panel) and south--north (lower panel) radial profiles of the stray light (lower set of curves) and of the F-corona (upper set of indistinguishable curves) over the [2004\,--\,2018] time interval in the case of the SOHO roll angle of 180$\degr$.}
\label{fig:Bf_SL_180}
\end{center}
\end{figure}

The photometric properties of the stray light are illustrated by radial profiles extracted along the east--west and south--north directions during the [2004\,--\,2018] time interval. 
The two cases of roll angles are displayed separately: 0$\degr$ in Figure~\ref{fig:Bf_SL_000} and 180$\degr$ in Figure~\ref{fig:Bf_SL_180}.
The profiles were taken at the same time of each year of the [2004\,--\,2018] time interval and piled up using different colors.
The ``F'' profiles are indistinguishable, as their temporal variations from year to year amount to at most a few percent.
The ``SL'' profiles exhibit substantial variability, but they tend to form sub-groups indicating the persistence of similar levels over several years and even of similar patterns, which implies that they are not noise.
These profiles rapidly decrease with increasing radial distance from the center so that, at the edge of the square \fov of LASCO-C2, it levels off at approximately $10^{-11}$\,\Bsun $\space$, that is a factor of $\approx$100 below the radiance of the F-corona in the equatorial direction and a factor of $\approx$60 in the polar direction, a quite remarkable achievement.

\begin{figure}[htpb!]
\begin{center}
\includegraphics[width=\textwidth]{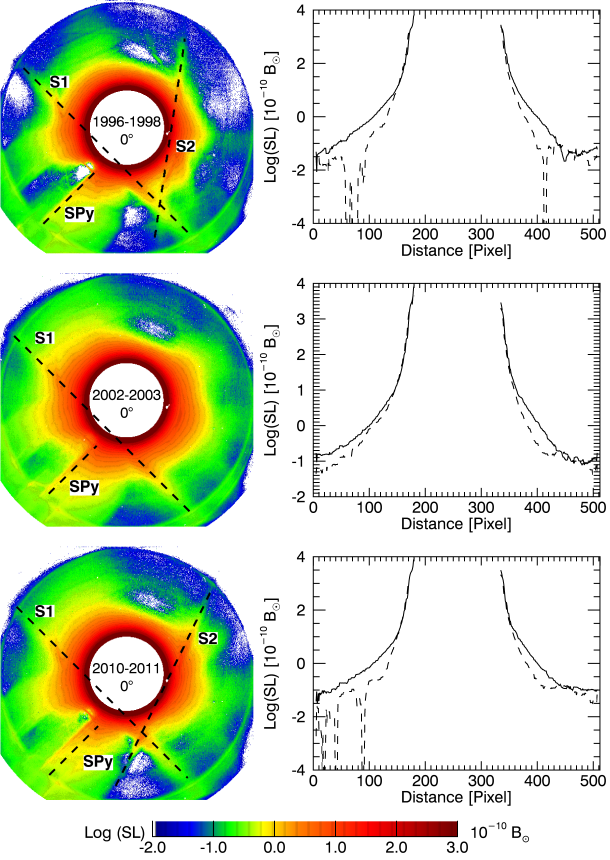}
\caption{Illustration of the main changes in the stray-light pattern of LASCO-C2 (left column) and radial profiles along the east-west (solid lines) and south--north (dashed lines) directions extracted from the images (right column).
These profiles were smoothed using a boxcar width of seven pixels for better legibility.
Whereas the S1 streak and the stray-light feature SPy connected to the pylon persisted throughout the mission, the S2 streak present at the beginning of the mission disappeared in February 2002.
The S3 streak appeared in 2010 and persisted thereafter.}
\label{fig:SL_Pattern}
\end{center} 
\end{figure}

The two-dimensional patterns of the stray light may be grossly characterized by several components: the narrow bright diffraction fringe, its associated broad wings extending to approximately two third of the \fovnospace, and different localized structures, some permanent like the radial reinforcement connected to one side of the pylon, some intermittent.
The most important changes that occurred during the 24 years are highlighted in Figure~\ref{fig:SL_Pattern}.
During the first years of operation, and until the accidental loss of SOHO in June 1998, the stray light remained constant and apart from the general pattern described above, exhibits two distinct streaks forming a V-shape, both being nearly tangent to the diffraction fringe.
The one oriented at 45$\degr$ (counterclockwise from the vertical direction) and labeled ``S1'' appears connected to the junction between the occulter and the pylon, although this may be fortuitous. 
The other one oriented vertically and labeled ``S2'' has no apparent origin.
Following the re-pointing of SOHO in early 1999, a slight increase of the overall level of stray light is revealed by our analysis but went unnoticed at that time.
In any case, as the diffraction fringe remained symmetric, a decision was made not to activate the mechanism controlling the centering of the inner occulter. 
Streak ``S1'' persisted, although becoming fuzzier, whereas ``S2'' completely disappeared in February 2002 and was later replaced by a new streak labeled ``S3'' at a different orientation ($\approx$-30$\degr$ from the vertical direction) and also tangent to the fringe, thus forming a new V-shape with ``S1''. 
``S3'' can be traced to the 2009\,--\,2010 image, but it became conspicuous thereafter, persisting until 2019. 
It is worth emphasizing at this stage that none of the components of the stray light patterns, in particular the streaks, can be attributed to residuals of the K-corona, implying that our procedure has been fully successful in eliminating any such traces.

\begin{figure}[htpb!]
\begin{center}
\includegraphics[width=\textwidth]{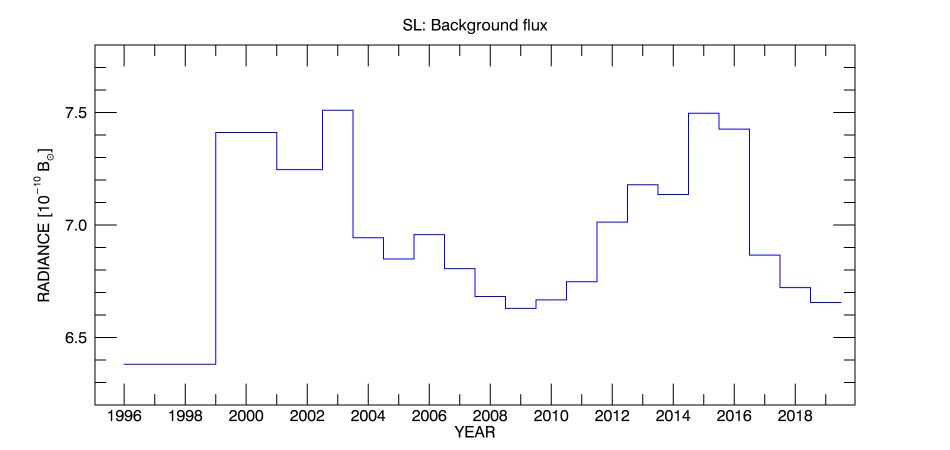}
\caption{Temporal variation of the background radiance of the stray-light images at roll angle of 0\deg integrated in the outer ring defined in the text and in Figure~\ref{fig:3rings}.
The temporal extent of each bin corresponds to the time interval of validity of the respective images.}
\label{fig:SL_Background}
\end{center} 
\end{figure}

The general background level of stray light was quantified by integrating its radiance in a ring that excludes the diffraction fringe. 
This ring encompasses the middle and outer rings introduced in Section~\ref{Sec:Stage1}, and thus extends radially from 86 to 250 pixels.
Figure~\ref{fig:SL_Background} displays its temporal variation in the case of the roll angle of 0${}^o$ since it extends over the whole interval of 24 years. 
It clearly reveals a correlation with solar activity with a modulation of $\approx$13\,\%, which implies that a small component of the stray light has its origin in scattered light from the innermost bright corona.

\newpage

\subsection{Stage 3: restoration of the F-Corona}
\label{Sec:Stage3}

With the stray-light images in hand,  we proceeded to restore the F-corona from the ``F+SL" set of images by subtracting from each one the appropriate ``SL" image.
As part of the final step of Stage 1, the ``F+SL" images were corrected for the varying Sun--SOHO distance according to the function $d^{-\scalebox{1.}{$\nu$}_{\scriptscriptstyle\rm F}}$.
This is irrelevant for the ``SL" images, which were thus corrected using the function $d^{({\scalebox{1.}{$\nu$}_{\scriptscriptstyle\rm F} - {\scalebox{1.}{$\nu$}_{\scriptscriptstyle\rm SL}}})}$ in order to restore their correct variation with $d$.
This procedure yielded a first time series of ``F'' images labeled ${S_3}$.
They are still modulated by the annual oscillation of SOHO around the symmetry plane of the zodiacal cloud, and this effect can be fitted point-by-point by a Fourier series as described below.

\begin{figure*}
\vspace{2cm}
\noindent
\centering
\includegraphics[height=\textwidth, width=0.43\textheight,angle=90]{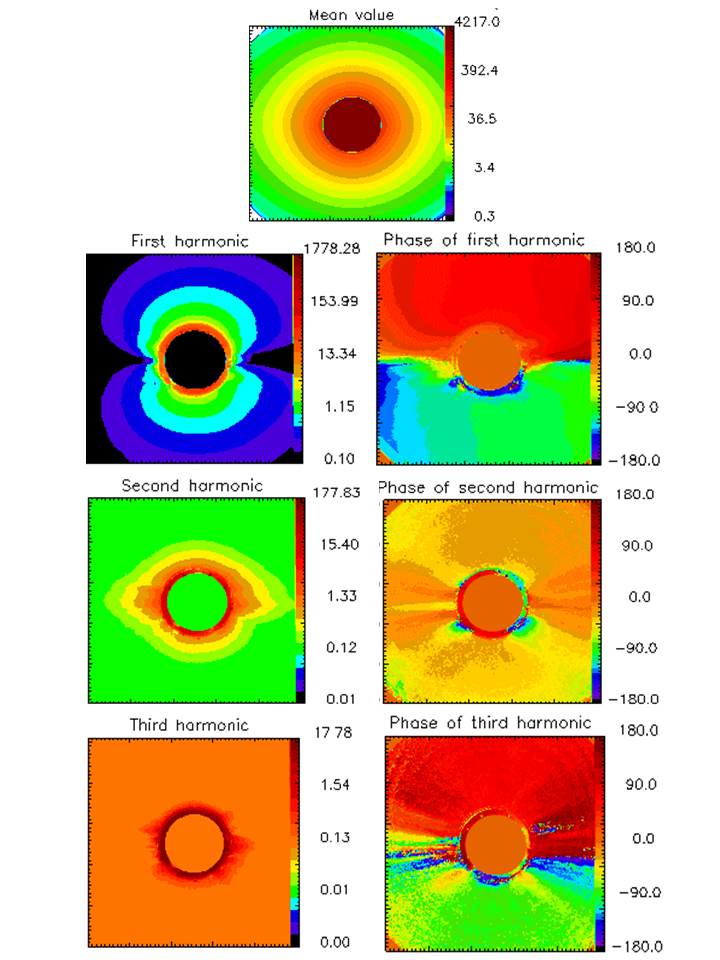}
\includegraphics[height=\textwidth, width=0.43\textheight,angle=90]{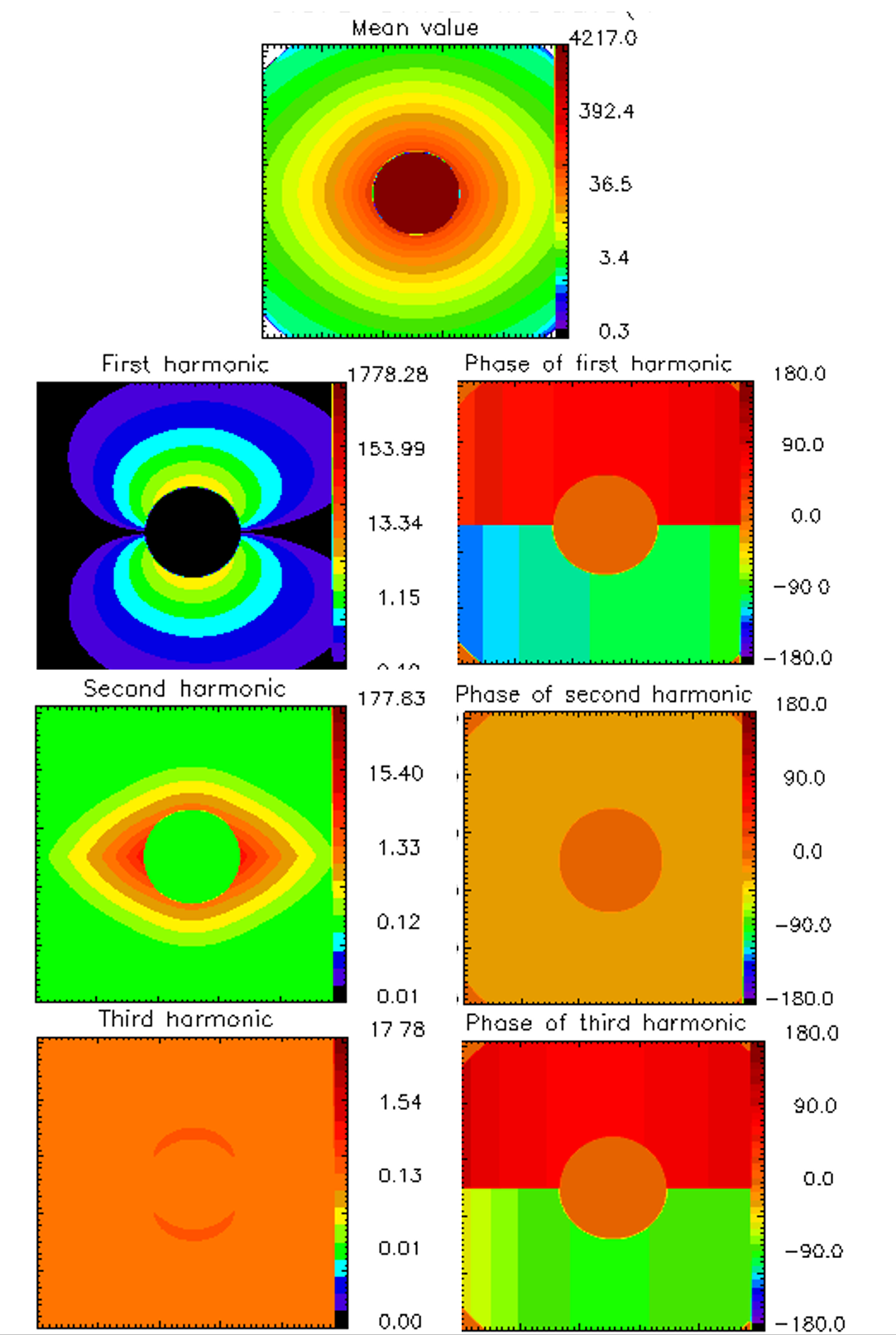}
\caption{Illustration of the Fourier analysis of a time series of ``F'' images (left block of images) and of the effect of filtering the coefficients (right block of images). 
In each block, the upper single image corresponds to the mean value.
Below, the modulus (left column) and phase (right column) of the coefficients are displayed for the first three harmonics.}
\label{fig:Harmonics}
\end{figure*}

We proceeded by structuring this ${S_3}$ series in blocks of two-year duration with an overlap of one year.
Each pixel of the coronal field was tracked along a given two-year interval, and each temporal profile was subjected to a Fourier analysis limited to the third harmonics.
This process led to the compression of the set of ``F'' images extending over a time interval of two years to a set of seven images: that of the mean value of the Fourier decomposition and those of the sine and cosine coefficient of the three harmonics.
As the harmonics' order increases, the images of the coefficients (ICs) become more and more sensitive to the local pixel-to-pixel noise in the ``F'' images. 
A bi-dimensional filtering was therefore applied to the six ICs after dividing them by the mean image to compensate for the strong radial gradient of the F-corona.
After filtering, each IC was individually restored by multiplying each filtered IC with the mean image. 
Ultimately, seven coefficients were used to fit the annual evolution of the F-corona: one for the mean, and 3$\times$2 for the three harmonics.
Figure~\ref{fig:Harmonics} displays an example such a set of seven images before and after filtering.

\begin{figure}[htpb!]
\begin{center}
\includegraphics[width=\textwidth]{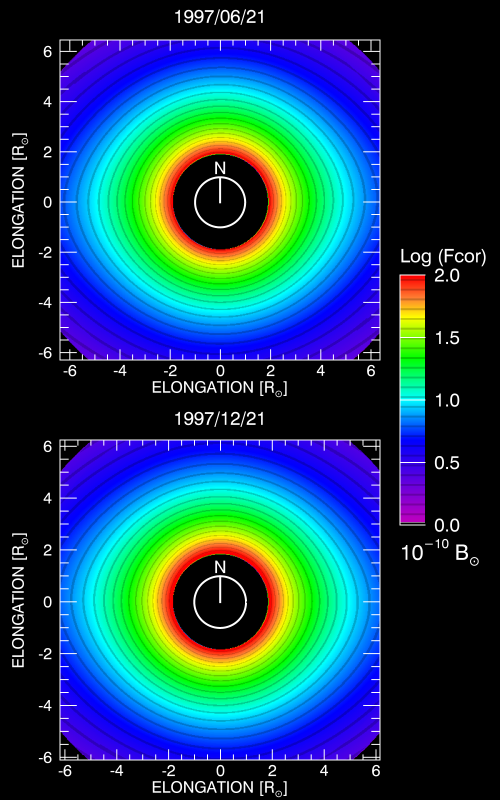}
\caption{Two images of the F-corona produced by the restoration procedure described in Section~\ref{Sec:Stage3} corresponding to the June (upper panel) and December (lower panel) 1997 nodes.
The latter image is brighter than the former reflecting the effect of the varying Sun--SOHO distance, which is nearly minimal in December and nearly maximal in June of each year.
Note the incidence on the elongation scale expressed in R$_\odot$.}
\label{fig:Final_F}
\end{center} 
\end{figure}

In order to mitigate the effect of the transition between the successive two-year time intervals, the validity of each set of Fourier coefficients was limited to a time interval of one year centered on each two-year interval, \ie from 1 July  of the first year to 30 June of the second one.
The restoration of the final images of the F-corona was performed on a daily basis forming the ``Fcor'' time series.
These images are in the format of 1024 $\times$ 1024 pixels, and their absolute radiance expressed in units of $10^{-10}$\,\Bsun $\space$ is given at the heliocentric distance of SOHO at the specified date, both reported in the image header.
Their orientations are such that North is always up. 
The vertical direction is aligned with the sky projected direction of the solar axis until 29 October 2010, and with the direction of the ecliptic (North) pole thereafter.
Examples are given in Figure~\ref{fig:Final_F} corresponding to two views, at the June and December nodes of 1997, when the Sun--SOHO distance was nearly maximum and minimum, respectively.

\subsection{Uncertainties Affecting the Restoration of the Stray Light and the F-Corona}
\label{Sec:Uncertain_SLF}

As argued in Article I, the error analysis of data resulting from complex operations is extremely difficult, and even elaborated approaches such as Monte-Carlo error propagation are  inapplicable. 
Ultimately, this led us to implement consistency checks and intercomparisons with several independent observations, prominently eclipse observations.
Let us however develop several arguments in favor of the high accuracy of our restorations.

First of all, the starting set of ``F+SL'' images was indirectly validated by the extensive intercomparison between simultaneous C2 and eclipses images of the total radiance $B$ and polarized radiance $pB$; the agreement was generally so compelling that it allowed us to construct several composite images ``eclipse + C2''.
For instance, in the case of the eclipse of 26 February 1998, the circular profiles extracted from $pB$ images obtained by C2 and by a team of the High Altitude Observatory (HAO) with their {\it Polarimetric Imager for Solar Eclipse 98} (POISE98) are in nearly perfect agreement except for a few local discrepancies that do not exceed 10\,\% (Figure 34 of Article I). 
The underlying assumption that the F-corona and the stray light are unpolarized, thus allowing their separation from the K-corona, is now further validated by a follow-on photopolarimetric analysis of the LASCO-C3 images in an article to be published shortly; it shows, in particular, that the assumption $p_{\scriptscriptstyle\rm F} = O$ holds at least out to $\approx$10\,R${}_\odot$, well beyond the C2 \fovnospace.

We showed above that, except for the diffraction fringe, it is quite difficult to disentangle the influence of the varying Sun--SOHO distance on the background stray light and on the F-corona.
However, the modulation introduced by this variation remains modest, $\approx$15\,\% in the middle ring and $\approx$8\,\% in the outer ring (Figure~\ref{fig:FluxesAll}), so that any error on this influence is certainly less than $\approx$2\,\%.

Regarding the restoration of the stray-light images, tracking the errors along the various steps of the procedure is almost impossible, but general trends may be delineated.
The diffraction fringe is so intense that it cannot be much affected by the underlying corona, as confirmed by its distinct variation with the Sun--SOHO distance.
The large-scale background stray light is more challenging as it may be contaminated by remnants of the K-corona and a fraction of the F-corona.
As already pointed out, the polarimetric separation had difficulties with bright K-corona structures, hence possible remnants,  but they were efficiently removed by the multiple averaging steps in our procedure; indeed, the stray-light images do not exhibit structures reminiscent of the K-corona.
Likewise, the large-scale spatial variations of the stray light are untypical of the smooth F-corona. 

Regarding the restoration of the F-corona, a relevant point comes from the comparison of the equatorial profiles of the Fcor images scaled at 1 AU with that of the ``K-L'' model elaborated by \cite{KoutchmyLamy1985}.
Indeed, the agreement is nearly perfect: the gradients are strictly similar and the absolute difference amounts to 6\,\%.
It should be understood that this does not mean that our Fcor images are off by 6\,\% since the ``K-L'' model was obtained by analyzing and synthesizing various observations, which have their own uncertainties. 
It is therefore a reasonable model, but its accuracy can certainly not be better than several percent.
However, the above agreement offers a very valuable mutual validation. 

\begin{figure}[htpb!]
\begin{center}
\includegraphics[width=\textwidth]{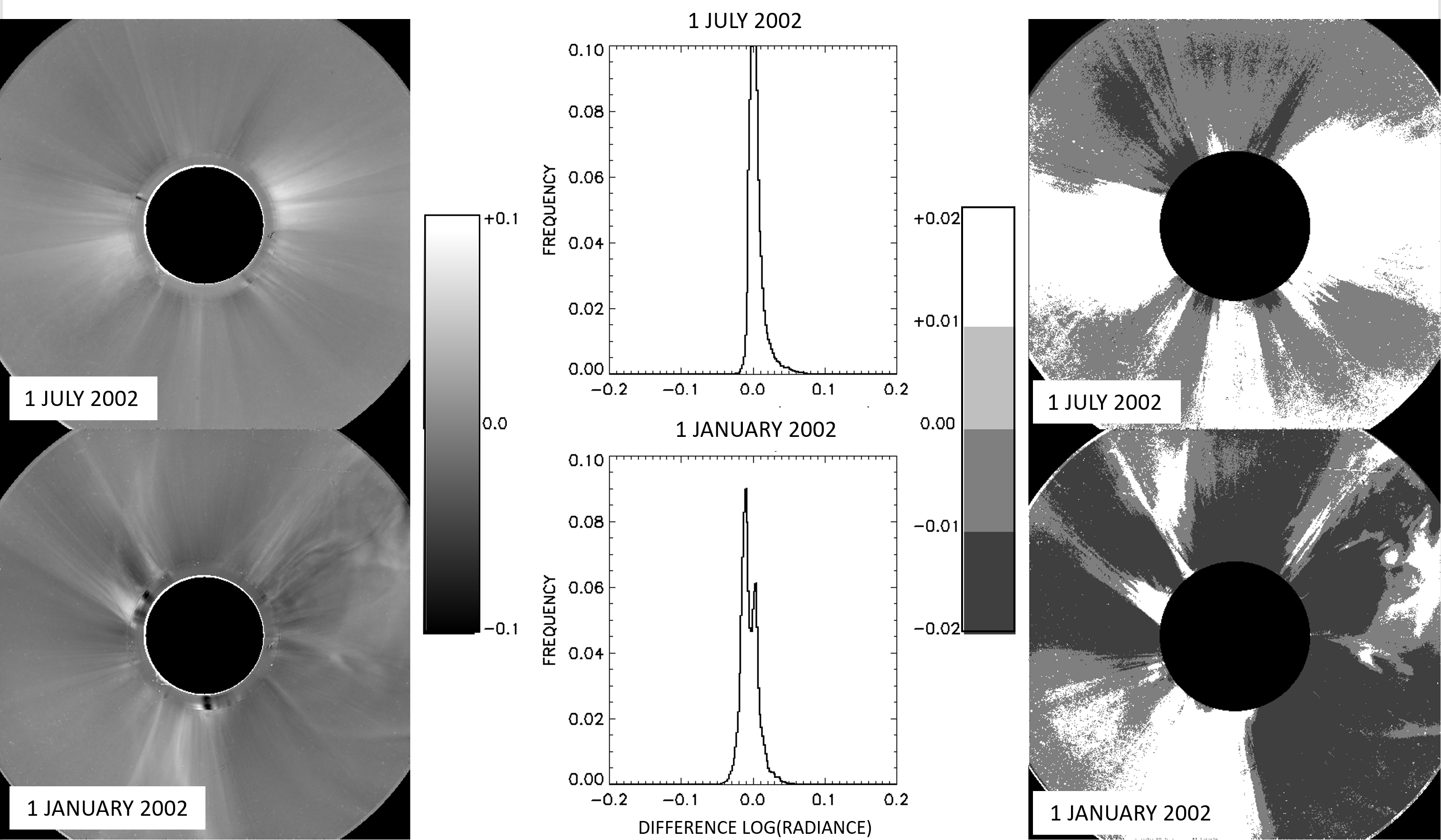}
\includegraphics[width=\textwidth]{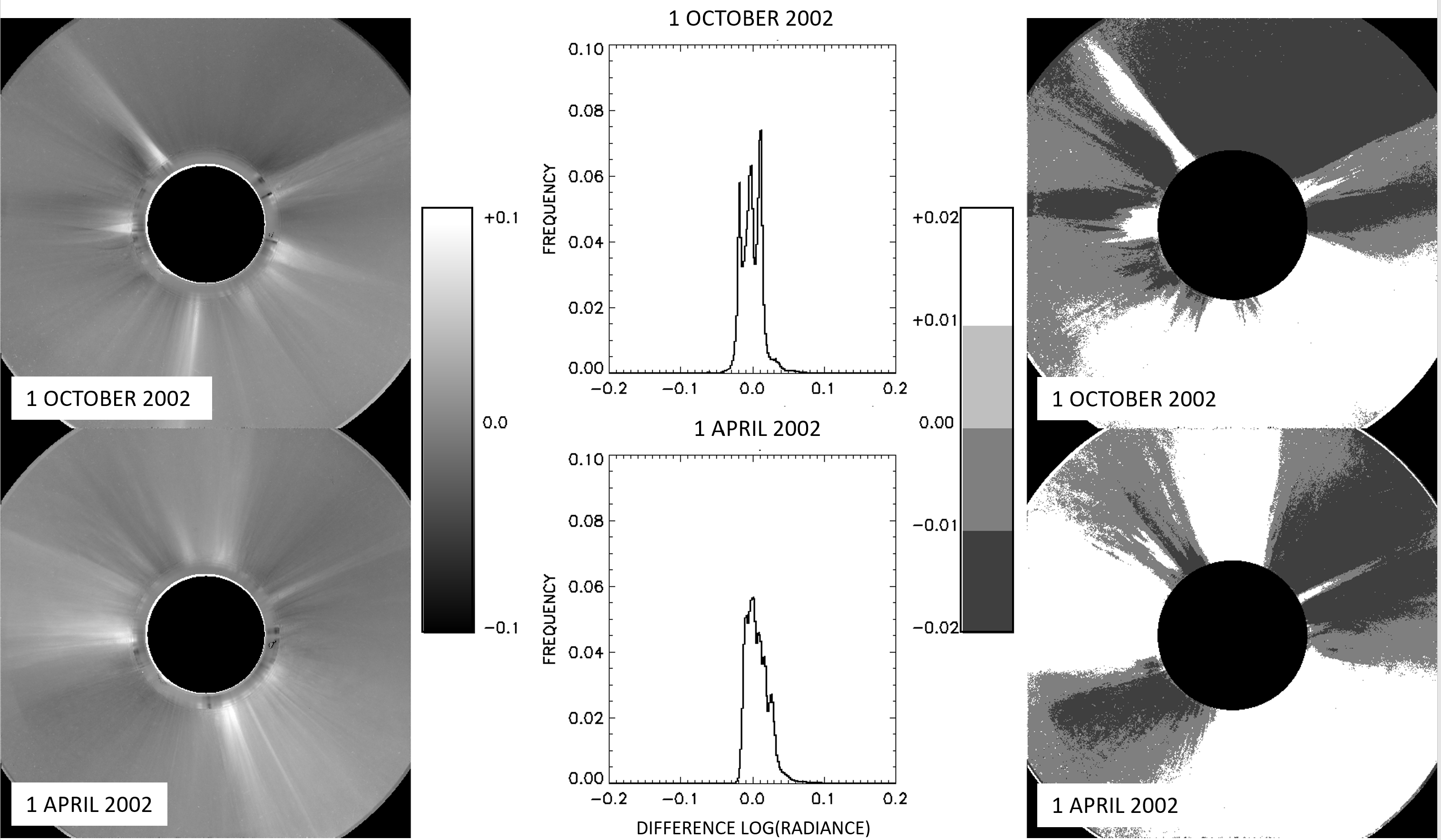}
\caption{Illustrations of the ratio between the original ``F+SL'' images and the restored ``Fcor+SL'' images combining the calculated stray light and the Fcor images at four dates of 2002.
In the upper two rows, the two dates correspond approximately to the times when the SOHO orbit crosses the plane of symmetry of the zodiacal cloud (the nodes).
In the lower two rows, the two dates correspond approximately to the times when SOHO is at locations in quadrature with the nodes.
Each panel is organized in three columns.
The left columns display the logarithm of the ratios in the range -0.1 to +0.1.
The right columns display regions where the logarithm of the ratios lies in four intervals defined by the gray bars.
The central columns display the frequency distributions (histograms) of the difference between the logarithm of the original and the restored images.} 
\label{fig:ratios}
\end{center}
\end{figure}

As a final consistency check, we recombined the calculated stray light and the Fcor images and compared them with the original ``F+SL'' images.
Four typical examples spanning the year 2002 (chosen at the peak activity of SC 23) are displayed in Figure~\ref{fig:ratios}.
The histograms of the difference between the logarithm of the original and the restored images clearly indicate that the relative errors remain in the narrow range of $\pm$2\,\%.
As an additional feature, the ratios conspicuously reveal the remnants or residuals of the K-corona resulting from the imperfect polarization separation as mentioned in Section~\ref{Sec:OVER}. 
Their contributions may be seen in the asymmetric tails of positive values in the above histograms and they do not exceed $\approx$5\,\%.

\newpage

\section{Restoration of the K-Corona High Resolution Images}
\label{Sec:KRestit} 

\subsection{Restoration Procedure}
\label{Sec:ResProc} 

With the images of the F-corona and of the stray light in hand, we are now in position to restore the K-corona from the routine, unpolarized, high-resolution images of 1024 $\times$ 1024 pixels by simple subtraction.
C2 took these images at a cadence of typically 67 images per day until February 2009 and 118 images per day thereafter, except during two major interruptions mentioned in Section~\ref{Sec:Oper}.
This leads to a data set of 626,234 images at the end of 2019.
As a matter of comparison, the respective numbers of the low resolution images of 512 $\times$ 512 pixels of the K-corona produced by the separation technique based on the analysis of polarization sequences \citep{Lamy2020} are typically one image per day until February 2009 and four images per day thereafter for a total of 21,303 images.
On the basis of the gain in spatial and temporal resolutions, it is expected that the subtraction approach will produce images of the K-corona suitable for improved tomographic reconstruction \citep{Frazin2010}.  
In addition, the polarimetric separation has inherent limitations resulting in the loss of some fraction of the K-corona signal. 
In other terms, this imperfect separation leaves some K-corona signal (\ie residuals) in our ``F+SL'' images as mentioned in Section~\ref{Sec:Stage3} and illustrated in Figure~\ref{fig:ratios}.

Our restoration procedure described below uses as input fully corrected and calibrated images generated by the routine pipeline processing developed at the Laboratoire d'Astrophysique de Marseille (formerly Laboratoire d'Astronomie Spatiale) as described in our past articles, notably \cite{Lamy2020}.
It corrects for all instrumental effects: bias, random errors in exposure times \citep{Llebaria2001}, missing telemetry blocks and impacts of cosmic rays \citep{Pagot2014}, and finally vignetting \citep{Llebaria2004}, and it applies a radiometric calibration to convert the raw data to units of $10^{-10}$ \Bsun $\space$.
These corrections and the calibration are periodically updated as appropriate, thanks to a continuous monitoring of the in-flight performance of C2.
The restoration procedure must take into account the temporal variation of the stray light and that of the F-corona as controlled by the heliocentric distance of SOHO, and with its roll.
During its development and test, we discovered an additional problem with the centering of the images to be subtracted, which turns out to be critical since the K-corona is only a small fraction of the F-corona. 
For instance, at 5\Rsun, the ratios K/F of the radiances are typically 0.15 and 0.03 in the equatorial and polar directions, respectively.
The centering problem arises from the fact that a slow drift of the center of the Sun as well as short-term fluctuations of typically a fraction of a pixel, but sometime up to a few pixels have been observed \citep{Llebaria2004}.
Therefore, the F-corona image built from the Fourier series Fcor as described in Section~\ref{Sec:Stage3} must be accurately re-centered before being subtracted. 
This was performed for each individual input image by considering its closest ``F+SL'' image, subtracting the ``SL'' image and performing a cross-correlation between the resulting ``F'' image and the daily Fcor images built from the Fourier series. 
 
The formalism to restore an image of the K-corona $K(i,j,t)$ at any given time $t$  from a corrected, high resolution, unpolarized image $I(i,j,t)$ may be conveniently expressed by the following equation:

\begin{eqnarray*}
K(i,j,t) & = & I(i,j,t) - C_{SL}(t){\times}SL(i,j,t)/\mathcal{V}(i,j) - {\rm Fcor}(i,j,t;i_0,j_0)
\end{eqnarray*}

\noindent using the following definitions:
\begin{itemize}
\item  $\mathcal{V}(i,j)$ is the C2 vignetting function,
\item  ${\rm Fcor}(i,j,t;i_0,j_0)$ is the Fcor image built from the Fourier series and centered at $(i_0,j_0)$, 
\item  $C_{\rm SL}(t)$ is a scaling coefficients applied to the stray-light images to account for the Sun--SOHO distance.
\end{itemize}

\begin{figure}[htpb!]
\begin{center}
\includegraphics[width=0.6\textwidth]{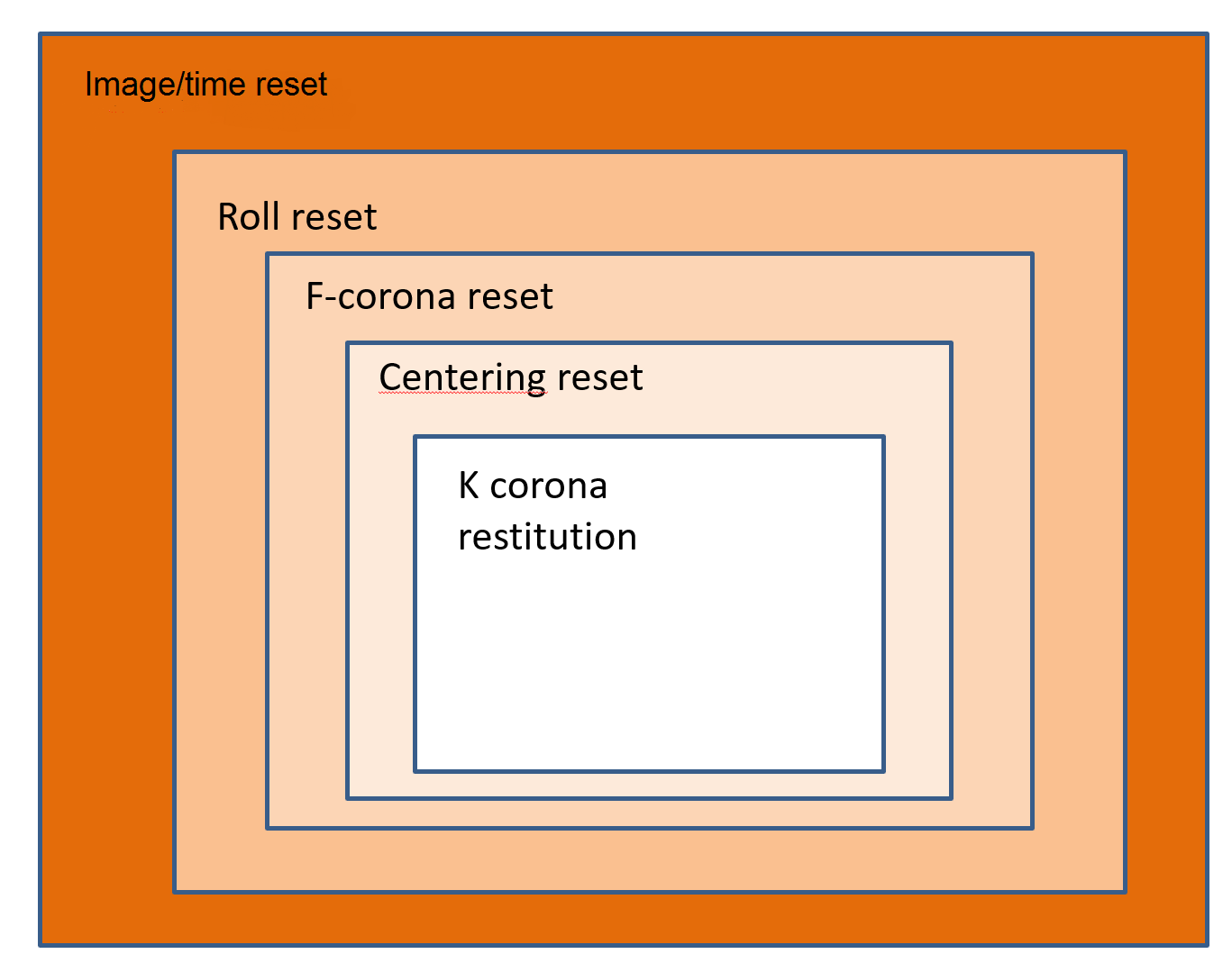}
\caption{Schematic of the embedded operations implemented to generate the high-resolution images of the K-corona from the routine LASCO-C2 images.} 
\label{fig:process3}
\end{center}
\end{figure}

The procedure summarized by this expression is applied on an annual basis to limit the number of images to be processed in a single run. 
The different time scales and the embedding of the different terms require a practical organization as a set of nested operations with specific event rates to reset in proper order the needed auxiliary images and parameters as schematically represented in Figure~\ref{fig:process3}.
The five nested operations perform the following tasks:

\begin{enumerate}[(a)]
\item Run the outermost loop over the input images controlled by the image index, which defines the absolute time $t$ (Modified Julian Date).
\item Check the roll status of SOHO so as to specify the ``SL'' image.
\item Retrieve the appropriate daily Fcor image and the closest ``F+SL'' image updated one to four times per day depending upon the image cadence.
\item Re-center this Fcor image.
\item Generate the K-corona image according to the above equation. 
\end{enumerate}

\subsection{Uncertainties Affecting the Restoration of the K-Corona}
\label{Sec:Uncertain_K}

These uncertainties may be analysis on the basis of the above equation that expresses the K-corona $K(i,j,t)$.
We first note that the vignetting function $\mathcal{V}(i,j)$ present in the second term does not introduce an additional uncertainty since it strictly compensates a first application of this function as explained in Section~\ref{Sec:Const_SL}.
Therefore, we are left with the uncertainties affecting the unpolarized images $I(i,j,t)$, which are controlled by the main operations applied to the raw images.
\begin{itemize}
	\item Random errors in exposure times are corrected using an elaborated equalization procedure, particularly efficient in the case of the high-cadence unpolarized images, ensuring an accuracy better than 0.1\,\% \citep{Llebaria2001}.
	\item The vignetting function was obtained from a detailed modeling of the optics of C2 ensuring an accuracy better than 0.3\,\% \citep{Llebaria2004}.
	\item The absolute photometric calibration makes use of thousand of measurement of around a hundred stars so that the C2 calibration factor is determined with an accuracy better than 1\,\% \citep{Lamy2020}.
\end{itemize}
It can further be underlined that these operations were applied as well to the polarized images resulting in polarized radiance images in excellent agreement with various eclipse observations, thus offering an independent verification of their high photometric quality.

\begin{figure}[htpb!]
\begin{center}
\includegraphics[width=0.6\textwidth]{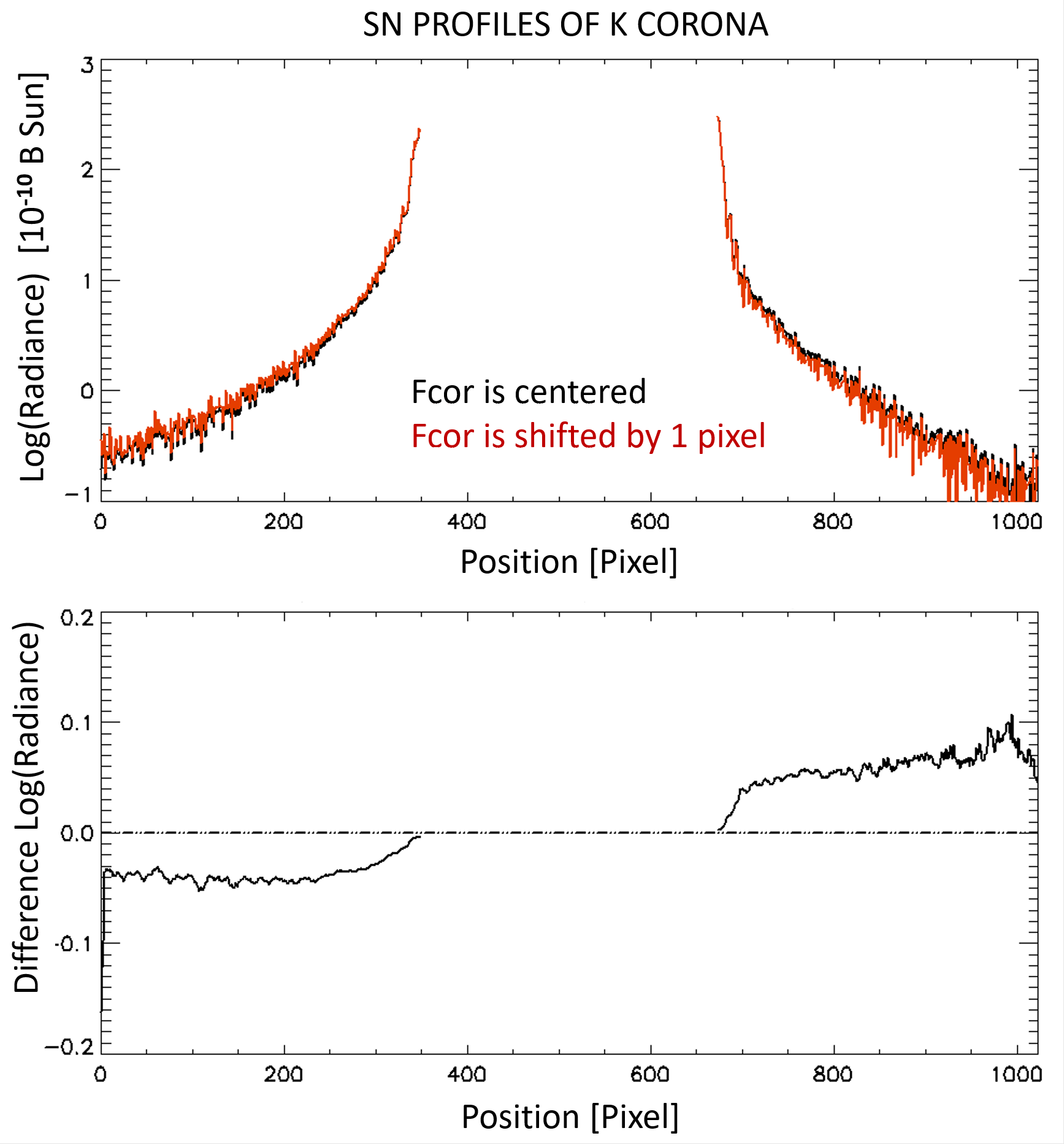}
\caption{Illustration of the effect of a an offset of one pixel in the centering of the Fcor image on the restoration of the K-corona.
The upper panel displays two North-South radiance profiles of the K-corona obtained with i) the properly centered Fcor image (black curve), and ii) the off-centered image by one pixel (red curve).
The lower panel displays the difference of the logarithm of the two radiance profiles.} 
\label{fig:shift}
\end{center}
\end{figure}

This error analysis of the terms of the above equation however misses a key aspect, which is unfortunately not amenable to a formal quantification: the critical centering of the Fcor images pointed out in Section~\ref{Sec:KRestit}, and which received special attention as described in this section.
Figure~\ref{fig:shift} illustrates the marked effect on the restoration of a K-corona image of an offset of one pixel on the centering of an Fcor of 1024 $\times$ 1024 pixels.
The resulting imbalance affecting the K-corona amounts to $\approx$-4\,\% along the South direction and up to $\approx$+8\,\% along the North direction.
We are confident that the correlation procedure implemented in Section~\ref{Sec:ResProc} ensures an optimal centering at the sub-pixel level. 
In summary, the above analysis indicates that the various uncertainties remain limited to at most a few percent, so that the global uncertainty affecting the high-resolution of the K-corona may be set at $\approx$10\,\%, with possible accidental exceptions resulting for instance from imperfect centering of the images. 

We very much regret the lack of images of the K-corona from ground-based, eclipse observations, thus precluding a direct comparison with our results as we performed for the $pB$ images in Article I.
As an alternative, we present in the next section relevant consistency checks by comparing various results on the radiance of the K-corona obtained by the two methods, subtraction (this work) and separation (Article I).

\begin{figure}[htpb!]
\begin{center}
\includegraphics[width=\textwidth]{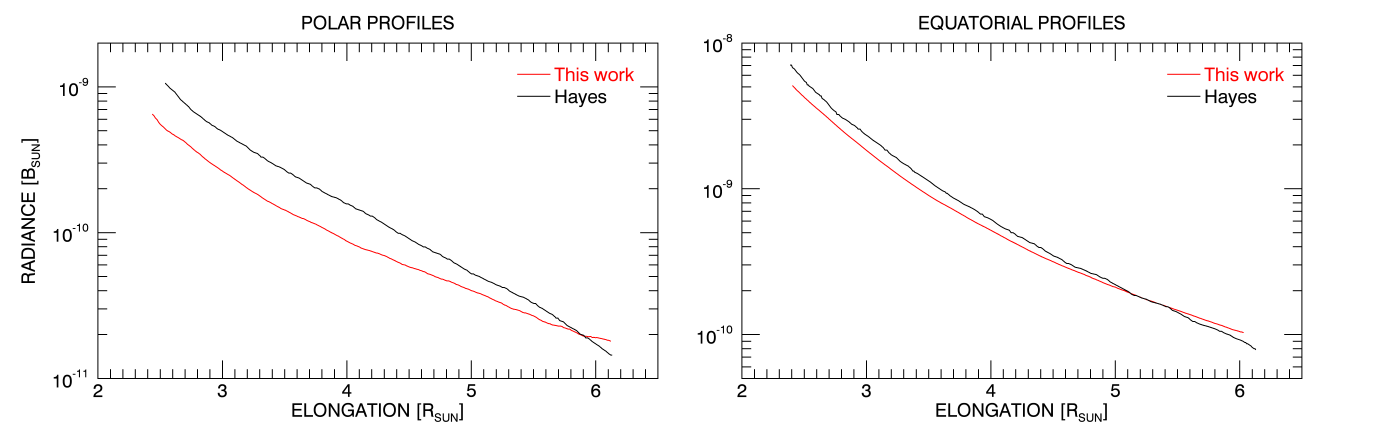}
\caption{Comparison of the equatorial and polar profiles of the radiance of the K-corona on 26 February 1998 determined by \cite{Hayes2001} (black curves) with those obtained with our restoration (red curves).} 
\label{fig:hayes}
\end{center}
\end{figure}

To conclude this section, we illustrate the importance of correctly assessing the background when retrieving the K-corona from the unpolarized LASCO-C2 images by comparing our results with those of \cite{Hayes2001} already mentioned in the Introduction.
We consider the same C2 image obtained on 26 February 1998 at 17:50 UT for which these authors provided two radial profiles, along the east equatorial and north polar directions as plotted in their Figure~3. 
Their ``total brightness'' denoted ``$B$'' is in fact the brightness of the K-corona obtained after subtracting a ``background minimum model as an approximation to the F-corona'', thus ignoring any stray light. 
We extracted similar profiles from our K-image after applying a Lee filter so as to obtain the same smoothness as the Hayes {\etal}'s profiles.
The differences plotted in Figure~\ref{fig:hayes} are readily understood as resulting from the different ``backgrounds'', being more pronounced in the case of the weaker polar profiles compared with the equatorial profiles.
In both cases, the gradients are nearly similar up to $\approx$4\,\Rsun and diverge beyond, consistent with the increasing role of the ``background'' with respect to the decreasing coronal signal.
This is clearly seen on Hayes {\etal}'s polar profile, which exhibits a suspect turnover at $\approx$5.5\,\Rsun.

\subsection{Comparison of the Restoration by Subtraction and by Separation}

\begin{figure}[htpb!]
\centering
\includegraphics[width=\textwidth]{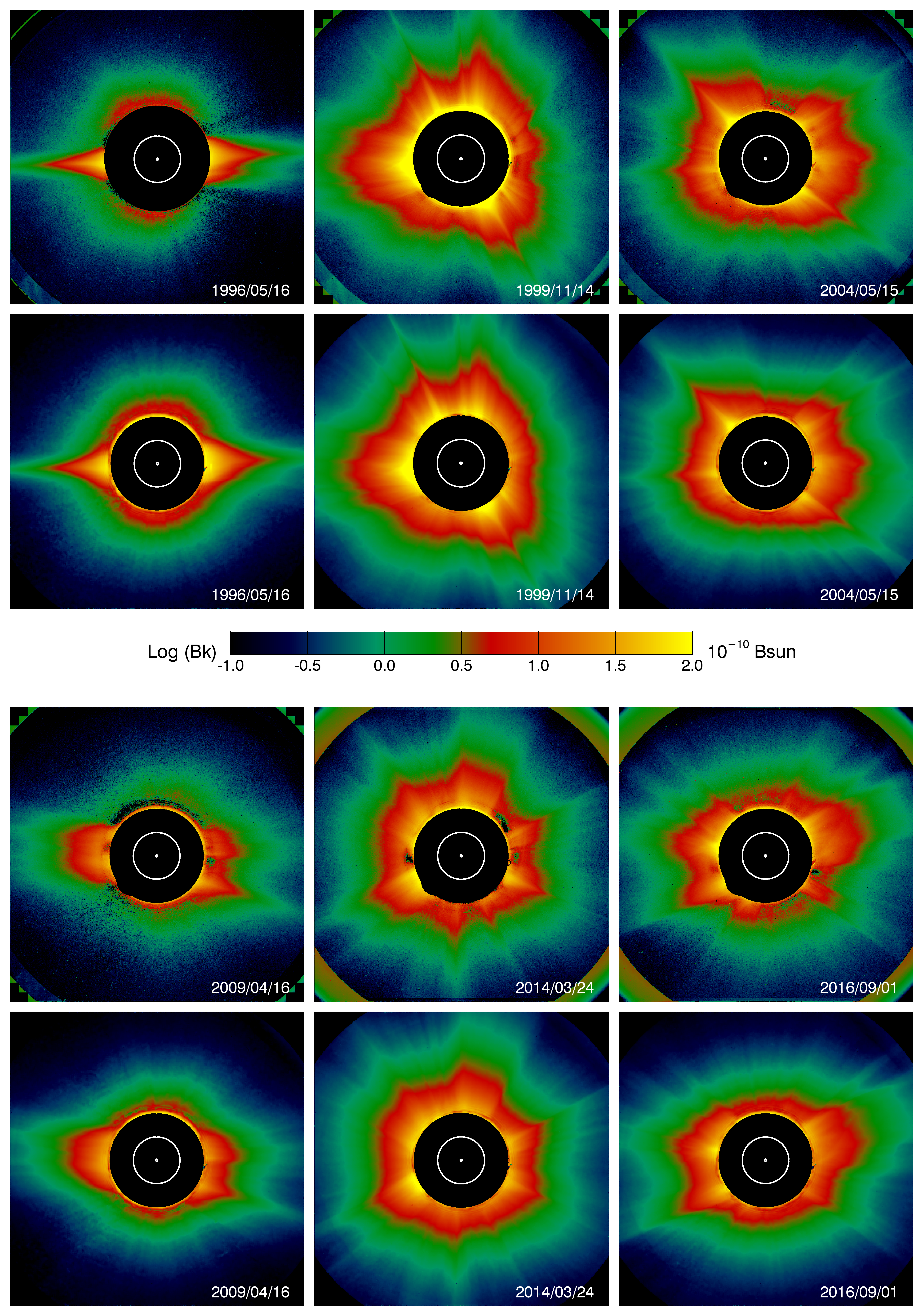}
\caption{Comparison of the maps of the radiance of the K-corona sampling SC 23 (upper group of six images) and SC 24 (lower group of six images), and obtained by subtraction (upper rows of each group) and by separation (lower rows of group).
The three columns correspond to three phases of activity of each cycle: minimum (left column), maximum (central column), and declining phases (right column). 
The dates are: 16 May 1996, 14 November 1999, and 15 May 2004 for SC 23 and 16 April 2009, 24 March 2014, and 1 November 2016 for SC 24. 
The white circles represent the solar disk and solar North is up.}
\label{fig:K1024_K512}
\end{figure}

\begin{figure}[htpb!]
\centering
\includegraphics[width=\textwidth]{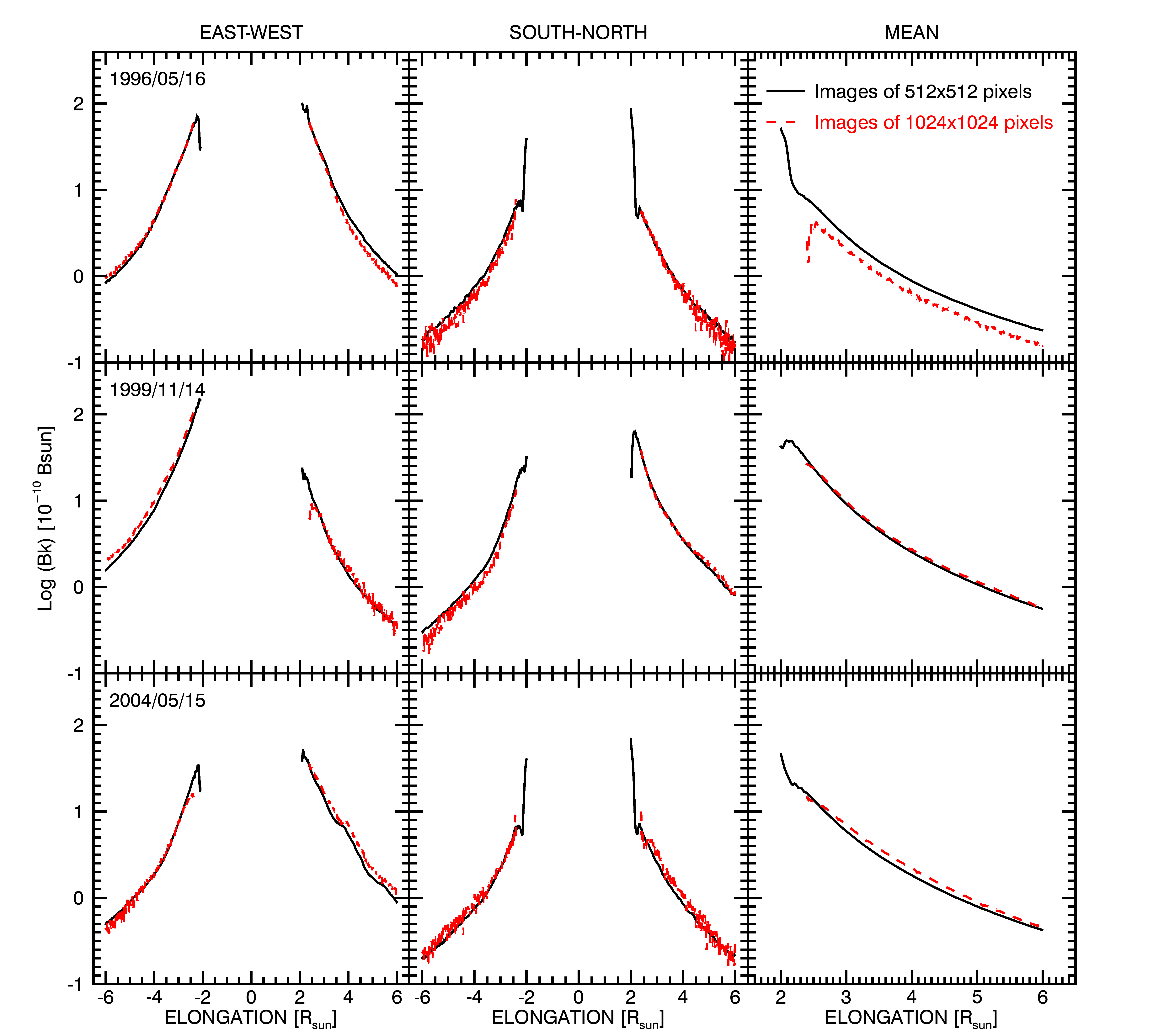}
\caption{Comparison of the profiles of the radiance of the K-corona obtained by subtraction (red-dashed lines) and by separation (black lines) constructed from the maps shown in the upper panel of Figure~\ref{fig:K1024_K512} and corresponding to three phases of SC 23: 16 May 1996, 14 November 1999, and 15 May 2004.
The radial profiles are displayed along the east--west (left column) and the south--north directions (central column), complemented by a mean profile (right column).}
\label{fig:K1024_K512_Prof1}
\end{figure}

\begin{figure}[htpb!]
\centering
\includegraphics[width=\textwidth]{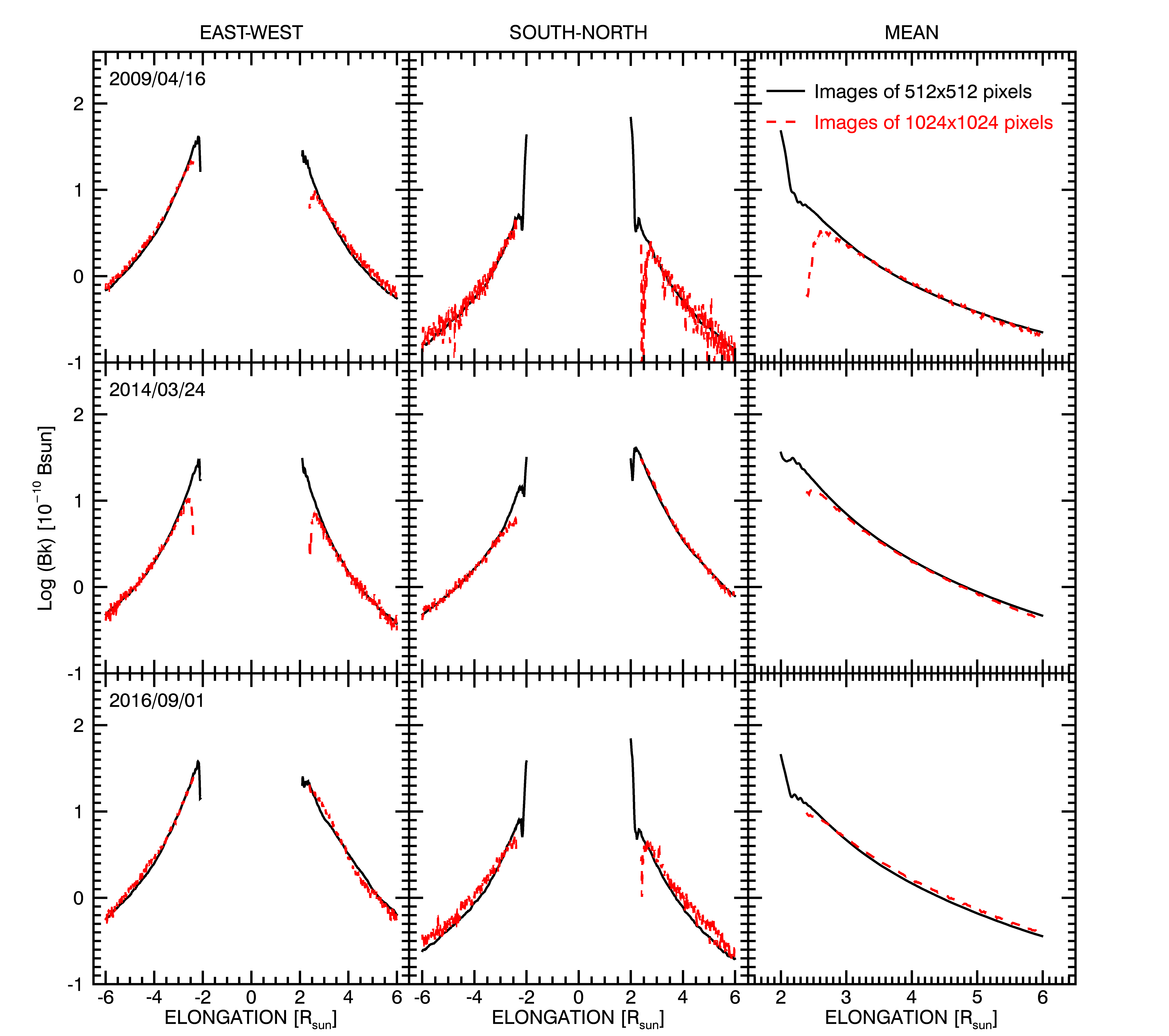}
\caption{Comparison of the profiles of the radiance of the K-corona obtained by subtraction (red lines) and by separation (black lines) constructed from the maps shown in the lower panel of Figure~\ref{fig:K1024_K512} and corresponding to three phases of SC 24: 16 April 2009, 24 March 2014, and 1 November 2016.
The radial profiles are displayed along the east--west (left column) and the south--north directions (central column), complemented by a mean profile (right column).}
\label{fig:K1024_K512_Prof2} 
\end{figure}

We now compare the radiance maps of the K-corona obtained by the two methods, subtraction and separation, limited to a few cases sampling SC 23 and 24 (Figure~\ref{fig:K1024_K512}).
This kind of display does not allow one to appreciate the gain in spatial resolution, but it does show the excellent agreement between the two methods.
A few discrepancies are present in images obtained by subtraction but always in the innermost region, close to the diffraction fringe.
This most likely reflects slight inaccuracies in the stray-light images and illustrates the difficulty of constructing these images to the finest detail.
A quantitative comparison is offered by Figures~\ref{fig:K1024_K512_Prof1} and \ref{fig:K1024_K512_Prof2} where equatorial, polar, and mean (averaged over 360$\degr$) profiles are compared.
Both the radial gradients throughout the \fov and the radiance levels are in excellent agreement with only a few slight discrepancies. 
It is interesting to note that the South--North profiles of the K-corona by subtraction are noisier than their counterparts by separation in the outer part of the \fov and during times of low or reduced activity, precisely when the radiance reaches levels below $10^{-10}$\,\Bsun $\space$.

\subsection{Temporal Evolution of the Radiance of the K-corona}

\begin{figure}[htpb!]
\centering
\noindent
\includegraphics[width=0.9\textwidth]{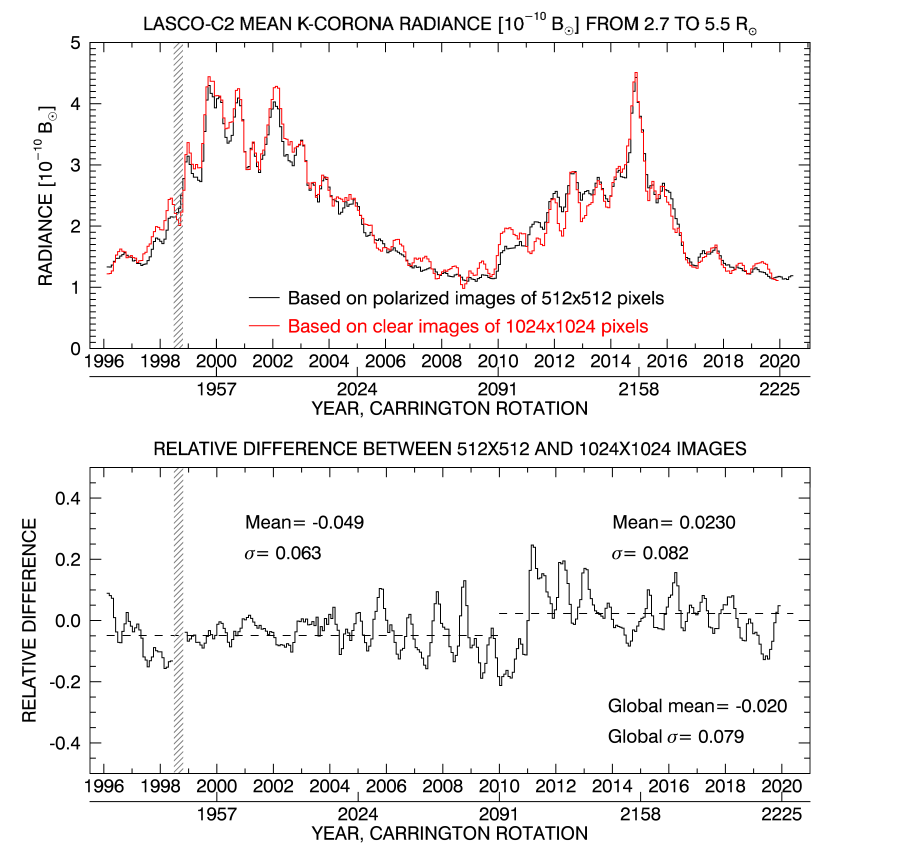}
\caption{Upper panel: temporal variations of the global radiance of the K-corona integrated from 2.7 to 5.5 \Rsun.
The two curves represent three-CR running means of the data averaged over the Carrington rotation.
The black curve corresponds to the case of the images obtained by separation (512 $\times$ 512 pixels), and the red curve to the case of the images obtained by subtraction (1024 $\times$ 1024 pixels)
Lower panel: relative difference between the two curves. 
The mean and standard deviation of the distributions are given for two time intervals and globally.}
\label{fig:evolution}
\end{figure}

We finally compare the long-term temporal evolutions of the radiance of the K-corona obtained by the two methods: subtraction and separation.
The radiance of the K-corona was globally integrated in an annular region extending from 2.7 to 5.5 R${}_\odot$ consistent with our past works, for instance \cite{Barlyaeva2015}.
Taking into account the varying SOHO--Sun distance, the size of the inner and outer radii expressed in pixels varied accordingly to precisely measure the same region of the solar corona throughout the 24 years.
The individual values of the integrated radiance were averaged over each Carrington rotation.
We applied a slight smoothing by taking three-CR running means to allow easier comparison between the two curves, and we calculated the relative difference between them as displayed in Figure~\ref{fig:evolution}.
The global agreement between the two evolutions is quite impressive, validating the inter-calibration of the two data sets as no scaling was introduced.
The two curves track each other closely with a high degree of correlation of the short-term variations, thus demonstrating the consistency of the two methods of restoration of the K-corona, except for larger mismatches during the ascending phase of SC 24, which probably reflect the difficulty of accurately restoring the stray light.
The relative difference (lower panel of Figure~\ref{fig:evolution}) has a global mean value of -2\,\%, but may be analyzed in terms of two regimes, incidentally corresponding to the two solar cycles although this may be fortuitous.
From 1996 to 2010, the relative difference fluctuates about a mean value of -5\,\% with a standard deviation of 6\,\% whereas beyond 2010, the corresponding values are 2.3\,\% and 8\,\%, respectively.
Still, they remain modest and fully demonstrate the validity of the two independent method of restoration of the K-corona.

\begin{figure}[htpb!]
\centering
\noindent
\includegraphics[width=0.9\textwidth]{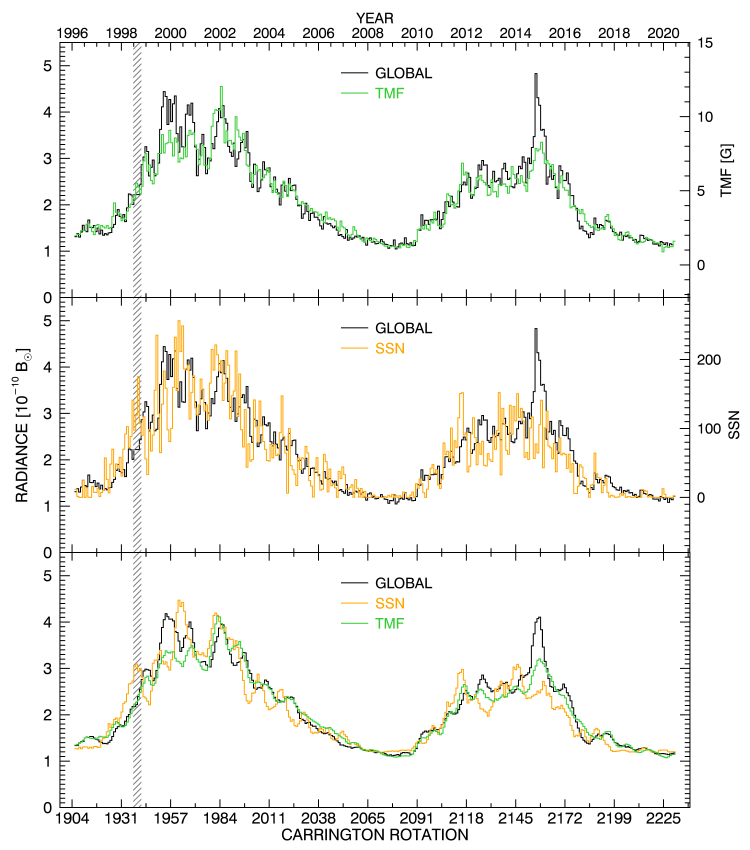}
\caption{Comparison of the temporal variation of the global radiance of the K-corona integrated from 2.7 to 5.5 \Rsun with those of the total magnetic field (TMF) and the sunspot number (SSN).
In the two upper panels, the time step is the Carrington rotation.
In the lower panel, we applied a smoothing by taking five-CR running means to allow an easier comparison between the three curves.}
\label{fig:Comparison_Bk_SSN_TMF}
\end{figure}

The main features of the temporal evolution of the K-corona have already been investigated by our team, notably its solar cycle variation until 2014.5 and the presence of mid-term quasi-periodicities by \cite{Barlyaeva2015}, the comparison between the minima of Solar Cycles 22/23 and 23/24 by \cite{Lamy2014}, and the anomalous surge at the end of 2014 by \cite{Lamy2017}.
We can now extend the comparison of minima to that of Solar Cycle 24/25 and make use of the data set resulting from the separation method since they are now available until mid-2020.
\cite{Barlyaeva2015} introduced several indices and proxies of solar activity to compare with the evolution of the radiance of the K-corona and also considered different latitudinal sectors.
For brevity, we presently limit the comparison of the global K-corona integrated between 2.7 and 5.5 R${}_\odot$ with the total magnetic field (TMF) and the sunspot number (SSN).
TMF was found to have the highest correlation with the K-corona whereas SSN has the lowest, but it is retained as the traditional indicator of solar (photospheric) activity.
The SSN data come from the WDC-SILSO data center\footnote{http://www.sidc.be/silso/datafiles}.
The TMF data are calculated from the Wilcox Solar Observatory photospheric field maps by Y.-M. Wang according to a method developed by \cite{Wang2003} and kindly made available to us.
The time sampling of the comparison is the Carrington rotation as allowed by the TMF data in order to consider variations at the highest possible frequency.
Figure~\ref{fig:Comparison_Bk_SSN_TMF} clearly confirms the conclusion of \cite{Barlyaeva2015} that the variations of the K-corona radiance show a remarkably high correlation with the TMF at all timescales.  
The TMF also tracks to some extent the surge experienced by the corona in late 2014, which is not the case of the SSN since it is a purely photospheric index.
Regarding the minimum of Solar Cycle 24/25, it is quite remarkable that the three data sets rigorously follow the same decreasing trend during the final years of the declining phase of SC 24, leveling off at the same low levels as those of the former ``anomalous'' SC 23/24 minimum. 
The subtle rising of the coronal radiance in early 2020, best seen on the smoothed profile (lower panel of Figure~\ref{fig:Comparison_Bk_SSN_TMF}), marks the transition to SC 25 in agreement with the SSN data.
In fact, correlating the two descending branches of SC 23 and 24 sets the duration of SC 24 at 11.0 years, precisely the value announced by the WDC-SILSO data center, \ie the canonical duration of a solar cycle.

\section{Conclusion}

In this article, we have presented the complex method developed to accurately determine the stray light and, in turn, the F-corona from the LASCO-C2 polarized images. 
This met our ultimate goal of the photometrically accurate restoration of the K-corona from the whole set of routine, high resolution images obtained over 24 Years [1996\,--\,2019], thus encompassing Solar Cycles 23 and 24.
This achievement would have been impossible without the availability of the polarization sequences and their prior in-depth analysis, which allowed us to perform the crucial, initial extraction of the ``F+SL'' images \citep{Lamy2020}.
It further benefited from the outstanding performances of C2, notably its sensitivity, its stability, and its very low level of stray light.
However, our method -- and this largely explains its complexity -- had to cope with adverse conditions such as the impact of the loss of SOHO, periodic rolls, and the switch of orientation from solar North to ecliptic North.
These changes are probably inevitable on long lasting space missions, but ultimately they did not hamper the photometric restoration over two solar cycles. 

It is worth pointing out the excellent agreement between the two sets of images of the K-corona, that produced by polarimetric separation and that presently produced by subtraction of the F-corona and stray light.
It however appears that the former method better handles the innermost region dominated by the bright diffraction fringe and more generally, the case of a faint K-corona characteristic of solar minima (Figure~\ref{fig:K1024_K512}).
This probably can be explained by the difficulty of performing an accurate restoration of the overwhelming stray light, which is the diffraction fringe, in this region.
In addition, the subtraction method induces noisier images of the K-corona images than the separation, as clearly shown by the radial profiles of Figures~\ref{fig:K1024_K512_Prof1} and \ref{fig:K1024_K512_Prof2}. 
This has a simple explanation, as the noise affecting the F-corona (primarily shot noise) is transferred to the K-corona since we used a noiseless model of the F-corona.
In principle, it could be eliminated or at least reduced by filtering as we implemented when we performed the comparison with the results of \cite{Hayes2001}.
The above remarks tend to favor the polarimetric method provided it can be performed at a high spatial resolution and at a high cadence, an improvement that would also benefit to the three-dimensional reconstruction of CMEs based on their polarization.
In that respect, advanced techniques as implemented at recent solar eclipses, such as micro-polarizer arrays affixed to the sensor (Burkepile et al.,2017; Vorobiev et al. 2017) should be considered for future space coronagraphs.
This should lead to more photometrically accurate maps of the K-corona as well as of the polarized radiance $pB$.
However, if one is interested in characterizing the F-corona, accurate subtraction of the stray light remains to be confronted. 

Concerning the implications for the K-corona observed over two solar cycles, we have confirmed that the temporal variation of the integrated radiance of the K-corona closely tracks the solar activity, and that it is highly correlated with the temporal variation of the total magnetic field.
We have further found that the behaviours of the integrated radiance during the last few years of the declining phases of Solar Cycles 23 and 24 are remarkably similar, reaching the same base level and leading to a duration of 11.0 years for the latter cycle, in agreement with the value from the sunspot number series.
Concerning the F-corona, we have revealed a long-term evolution of its radiance that clearly does not follow the solar cycle and this suggests a secular variation intrinsic to the inner zodiacal cloud.
This will be further explored in a forthcoming article that will combine the observations by both C2 and C3. 

\begin{figure*}
\vspace{0.5cm}
\noindent
\centering
\includegraphics[height=\textwidth, width=0.91\textheight,angle=90]{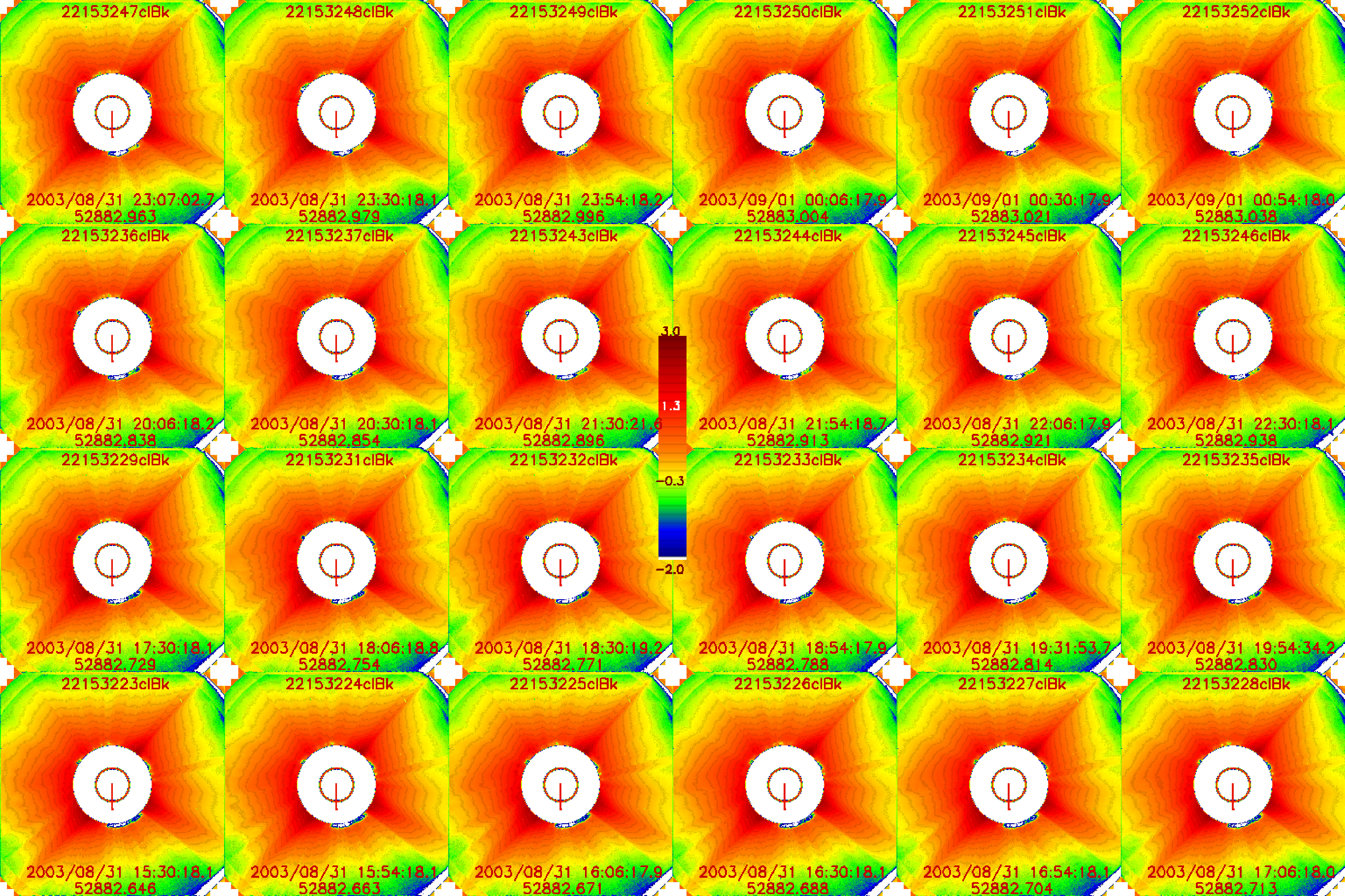}
\caption{An example of a page displaying 24 images of the reconstructed K-corona from the LASCO-C2 Legacy Archive.}
\label{fig:page}
\end{figure*}

The complete data set of the reconstructed high resolution images of the K-corona (FITS files) as well as the set of daily ``Fcor'' images are available from the LASCO-C2 Legacy Archive\footnote{\url{http://idoc-lasco-c2-archive.ias.u-psud.fr}} hosted at the Integrated Data and Operation Center (formerly MEDOC) of the Institut d'Astrophysique Spatiale.
The K-corona images may be conveniently visualized thanks to a set of pages, each one regrouping 24 compressed images at the format of 256$\times$256 pixels.
Each image is labeled with the original name of the source image and with the time expressed in ``yyyy/mm/dd hh:mm:ss'' and in MJD--50,000, where MJD is the Modified Julian Date.
The radiance of the K-corona is displayed in units of $10^{-10}$ \, \Bsun $\space$ using a logarithmic scale. 
Each group of 24 consecutive images arranged in a 6 $\times$ 4 format in lexicographical order from bottom to top is presented in a page. 
Each page is identified by its name which incorporates the date (yyyymmdd) and time T (hhmm) of the last image, followed by the date expressed in MJD (without the decimal point), for example ``map$\_$20030905$\_$1030$\_$528874''.
A color bar is placed at the center of each page. 
Figure~\ref{fig:page} displays an example of such a page.
The set of pages covering a month of observation (typically 120 pages) is regrouped in a document in PDF format. 
For the first 24 years of operation, there are 284 documents since four months are missing due to the loss of SOHO.
These PDF documents are accessed via a dedicated table where the entries are year and month. 

\vspace{\baselineskip} 
\noindent
{\footnotesize\bf Acknowledgments}
We thank Y.-M. Wang for providing the Total Magnetic Field (TMF) data.
The LASCO-C2 project at the Laboratoire d'Astrophysique de Marseille and the Laboratoire Atmosph\`eres, Milieux et Observations Spatiales is funded by the Centre National d'Etudes Spatiales (CNES).
LASCO was built by a consortium of the Naval Research Laboratory, USA, the Laboratoire d'Astrophysique de Marseille (formerly Laboratoire d'Astronomie Spatiale), France, the Max-Planck-Institut f\"ur Sonnensystemforschung (formerly Max Planck Institute f\"ur Aeronomie), Germany, and the School of Physics and Astronomy, University of Birmingham, UK.
SOHO is a project of international cooperation between ESA and NASA.

\vspace{\baselineskip} 
\noindent
{\footnotesize\bf Disclosure of Potential Conflicts of Interest} The authors declare that they have no conflicts of interest.

\bibliographystyle{spr-mp-sola}
\bibliography{Restoration-K-F_Biblio}                
\nocite{*}




\end{article}

\end{document}